\documentclass[preprint]{aastex}

\usepackage{graphicx}

\usepackage{natbib}
\usepackage{amsmath}

\def\<<{{\ll}}
\def\>>{{\gg}}
\def\spose#1{\hbox to 0pt{#1\hss}}
\def\ltwig{\mathrel{\spose{\lower 3pt\hbox{$\mathchar"218$}}
     \raise 2.0pt\hbox{$\mathchar"13C$}}}
\def\gtwig{\mathrel{\spose{\lower 3pt\hbox{$\mathchar"218$}}
     \raise 2.0pt\hbox{$\mathchar"13E$}}}

\def\beq{\begin{equation}}
\def\eeq{\end{equation}}

\def\=={\equiv}
\def\nuhat{{\bf \bar \nu}}
\def\vmbar{{{\hat{V}}_m}}
\def\omegagbar{{{\hat{\Omega}}_\gamma}}
\def\omeganbar{{{\hat{\Omega}}_\nu}}
\def\amax{{A_{max}}}
\def\scriptt{{\mathcal{H}}}
\def\scriptr{{\mathcal{R}}}

\def\lnu{{L_\nu}}
\def\lgamma{{L_\gamma}}

\def\field{{\mathcal{E}}}

\def\asphere{{A_o}}
\def\edim{{  $\mu$V\ m$^{-1}$MHz$^{-1}$  }}
\def\mm{{$^{-1}$}}

\def\tpar{t_{\parallel}}
\def\rpar{r_{\parallel}}
\def\Tpar{{\hat{t}_{\parallel}}}

\begin{document}

\title{Analytic Aperture Calculation and Scaling Laws for Radio Detection of 
Lunar-Target UHE
Neutrinos}

\author{K. G. Gayley, R.L. Mutel, T. R. Jaeger}
\affil{University of Iowa, Iowa City, IA 52242}

\begin{abstract}
We derive analytic expressions, and approximate them
in closed form, for the
effective detection aperture for Cerenkov radio emission
from ultra-high-energy neutrinos striking the Moon. 
The resulting apertures are in good agreement 
with recent Monte Carlo simulations and
support the conclusion of \cite{James:2009} that 
neutrino 
flux upper limits derived from the GLUE search \citep{Gorham:2004} 
were too low by an order of magnitude.
We also use our analytic expressions to derive 
scaling laws for the 
aperture as a function of observational and lunar parameters. 
We find that at low frequencies downward-directed neutrinos always dominate, but at higher frequencies, the contribution from upward-directed neutrinos becomes increasingly important, especially at lower neutrino energies.
Detecting neutrinos from Earth near the GZK regime
will likely require radio telescope arrays with extremely
large collecting area ($A_e\sim 10^6$ m$^2$) and hundreds of hours exposure time. Higher energy neutrinos
are most easily detected using lower frequencies.
Lunar surface roughness is a decisive factor for obtaining detections at
higher frequencies ($\nu \gtwig$ 300 MHz) and higher energies
($E \gtwig 10^{21}$ eV).

\end{abstract}

\section{Introduction}

The ubiquitous presence of ultra-high-energy 
(UHE, $E_{\nu}>10^{18}$ eV) cosmic rays suggests
the existence of an equally ubiquitous and similarly 
high-energy cosmic neutrino
population, either as a result of the various mechanisms for generating
muon neutrinos from charged pion decay in the vicinity
of the cosmic-ray acceleration region 
\citep{Bahcall:2001},
or
during interactions with the
cosmic background radiation
\citep[GZK effect,][]{
Greisen:1966, Zatsepin:1966}.
One effort to detect such neutrinos involves radio
observations of the expected Cerenkov burst emission when these neutrinos
interact with the Moon. Several experiments have already attempted to detect this signal 
\citep[e.g.,][]{Hankins:1996,Gorham:2004,Beresnyak:2005,Buitink:2008}
all with null results to date.
But even null results can be converted into 
useful upper bounds for the UHE neutrino flux, provided that 
the effective aperture, the area times the solid
angle through which incident neutrinos are detectable, is known.

For future experiments, it is clear that 
the larger the aperture, the greater will be the possibility
of achieving a detection, or the more decisively
constraining will be the inferred upper limit.
Of particular note is the fact that all the experiments so far
have suffered from particularly weak coverage of the energy domain near
the GZK cutoff near $10^{19.6}$ eV, the region of perhaps
the highest cosmological interest, and also possibly a local peak
in the neutrino spectrum.
The poor coverage is largely due to inherent limitations on radio observations,
but it may also in part be due to the difficulty in achieving
optimal tailoring of the observing parameters.
Having access to closed-form expressions of general validity
would yield not only a convenient means for calculating the aperture, it
would also assist in such
optimization of experimental design, for
penetrating deeper into this hitherto uncharted neutrino regime.

Despite their potential value,
no such closed-form expressions are currently available in the literature,
and so the primary purpose of this paper is to provide such expressions.
We then apply them to the question of how to optimize for detection of GZK
and other UHE
neutrinos by looking for radio Cerenkov signals from the Moon.
The approach is general enough for future modifications to accommodate
other
types of experiments, such as terrestrial ice 
sheets seen from airborne balloons,
or the Moon seen from lunar orbit.

To account for the neutrino properties and their
detectability in closed-form aperture expressions, 
we rely on previous determinations of the basic
attributes of radio Cerenkov emission from ultra-high-energy
(UHE) hadronic showers \citep[e.g.,][]{James:2009}
and reviews of the basic physics of the Askar'yan effect,
whereby neutrinos generate charge excesses in hadronic showers 
\citep[e.g.,][]{Alvarez-Muniz:2001}.
Here we will not comment on these physical
processes, but merely quote the results
from the literature as we apply them to our specific problem.

In this paper we assume that all contributing
neutrino showers occur in the lunar regolith
near the surface, with a fixed index of refraction of
$n_r = 1.73$ \citep{James:2009}, so we ignore 
inhomogeneities such as the sub-regolith.
\cite{James:2009} have evaluated a sub-regolith contribution and 
find that it can only significantly increase the aperture for
Earthbound detection at 
very high neutrino energies 
($E \gtwig 10^{22}$ eV), because only such high-energy neutrinos
produce sufficiently strong fields to be above the minimum detectable level at the surface. Of course, experiments from a closer distance, such as lunar orbiters 
like LORD \citep{Gusev:2006}, could receive significant contributions
from the sub-regolith, and ignoring gradients in the lunar material
introduces potential errors but is certainly a greatly simplifying assumption.

\section{General form of the aperture calculation}

Our goal is to determine the rate of detection of 
hadronic showers initiated
in the Moon by
the capture of a high-energy cosmic neutrino, via
the detection of the resulting radio-frequency
electric fields propagating from
the Moon.
One way to conceptualize this detection rate is in terms of an
aperture size, which when multiplied by
the incident neutrino flux in each energy
bin of interest, gives the detection rate in that bin.
Since the diffuse neutrino background flux
is scaled per area and per solid angle, the 
aperture will be in units of an area times a solid angle \citep{Williams:2004}.

Our approach for analytically specifying this aperture begins with
identifying the maximum possible
aperture $\amax$ that an amount of mass equivalent to the lunar
mass could possibly achieve,
assuming that each neutrino interacting with that material
is detected exactly once.
Note this would actually require the lunar material be unphysically 
spread out, among other impossible requirements.
Thus it is merely a starting point, and it follows clearly that
\beq
\label{emass}
\amax(E) = 4 \pi M \kappa(E)
\eeq
for mass $M$ (of the Moon) exhibiting 
a cross section per gram $\kappa(E)$ for
initiating charged particle showers for neutrinos at energy $E$, incident
over the full 4$\pi$ steradians of sky.

Our approach is then to reduce the aperture
by eliminating events that are blocked from occuring by virtue of
prior absorption of the neutrino elsewhere in
the spherical Moon, and then reduce it further by requiring that the
events yield radio signals above
the detection threshold of the specified telescope system along
some ray that intersects the detector.
These reductions are
substantial for three reasons:
ultra-high-energy (UHE) neutrinos
are drastically truncated by the opaqueness of the Moon, 
total internal reflection at the lunar surface
reduces the rays
that successfully cross the falling refractive index at this 
boundary, and rays from depth in the
Moon are significantly attenuated by lunar radio absorption.
The remainder of this paper is devoted to quantifying the aperture reductions
stemming from these three effects.

\subsection{The phase-space partition}

In order to include these 
corrections, we first subdivide $\amax$ by multiplying it
by the {\it fractional} (normalized
to unity) phase-space volumes
of all the processes that contribute
to $\amax$,
and then weight
each subdivision by an efficiency fraction, or probability, 
by which that phase-space component
contributes to the {\it observable} aperture.
Hence,  the detection rate for a neutrino flux distribution
$I(E,{\nuhat})$, where $I(E,{\nuhat})$ is 
per energy bin at energy $E$ and per solid angle along direction
$\nuhat$ and per target area, is
\beq
\label{rate-formal}
\scriptr \ = \ \int d \vmbar \int d \omegagbar
\int  d \omeganbar \int  dE \ 
\amax(E) \ I(E,{\nuhat}) e^{-\tau_\nu} \ {\scriptt_D}\
{\scriptt_R} \  \ ,
\eeq
where 
the phase space consists of three dimensions of lunar volume
$d \vmbar$ that account for all possible shower
locations inside the Moon,
two dimensions in $d \omegagbar$ that
account for the possible directions of the
rays along which the electric field of the shower can propagate (after
leaving the Moon),
and two dimensions in $d \omeganbar$, which account for 
the possible directions of the incident
neutrino.
The weighting function 
$e^{-\tau_\nu}$ accounts for the penetrating fraction of neutrinos which 
reach the phase-space contribution element
under consideration (the first correction mentioned above),
the Heaviside step function $\scriptt_R$ selects the outgoing radial
rays from $d\omegagbar$ that do not totally internally reflect at
the lunar surface (the second correction), and $\scriptt_D$ further selects 
the rays from $d\omegagbar$
that are bright enough to detect (the third correction).

\subsection{Surface roughness effects}

To determine whether or not a particular ray will escape total internal
reflection at the lunar surface, it is necessary to know the angle that
the ray meets the surface, which can be altered by lunar surface roughness
on the wavelength scale or larger.
\citet{Shepard:1995} analyzed radar reflections from the lunar surface
and found that the surface irregularities are self-similar  
(fractal), and the root-mean-square
roughness angle, which we convert to a Gaussian halfwidth by
multiplying by $\sqrt{2}$, can be parameterized as
\beq
\sigma_o(\lambda) = 
\sqrt{2} \ \tan^{-1}\left(0.29 \ {\lambda}^{-0.22}\right) = \sqrt{2}\ {\rm tan}^{-1}(0.14\ \nu^{0.22})
\label{sigma-surface}
\eeq
where $\lambda$ is the spatial scale in cm, $\nu$ in GHz,  and $\sigma_o$ is in radians.

\begin{figure}[h]
\begin{center}
\includegraphics[width = 6in]{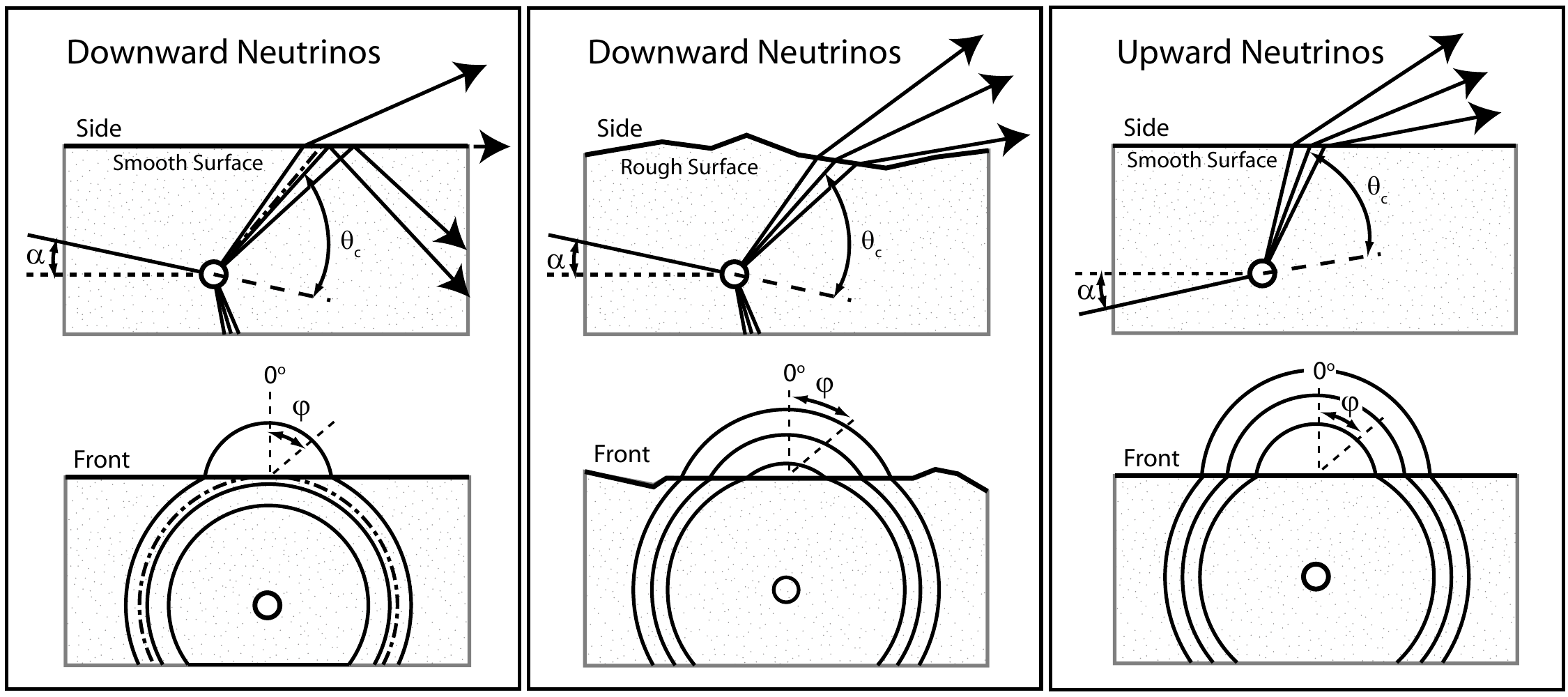}
\caption{(Left panel) Side and face-on views of Cerenkov cone 
ray paths escaping the lunar surface for downward-directed 
neutrinos, assuming a smooth surface. 
(Middle) Same, but including surface roughness. 
Note that a larger fraction of the Cerenkov cone escapes. 
(Right) Same, but for upward-going neutrinos. }
\label{Fig:Geometry}
\end{center}
\end{figure}

Even though
the surface may be tilted symmetrically
in any direction, the way surface roughness alters the local refraction angle
may have an important impact because of the
nonlinearity of the Heaviside step functions.
This nonlinearity can be particularly important for ``downward''
neutrinos,
which are the most likely to produce total internally reflected
rays becasue of their downward-pointing Cerenkov cones.
To illustrate this,
Fig.~\ref{Fig:Geometry} depicts schematic ray paths of the 
escaping Cerenkov cone both with and without surface roughness. 
In general, favorable surface tilts increase the number of rays escaping from 
downward-directed neutrinos more than the unfavorable tilts reduce it,
because there may already have been little or no signal prior to the
inclusion of surface tilt (and note that once a signal contribution
becomes zero it never becomes 
{\it negative}, so unfavorable tilts come with little penalty).
The enhancement becomes
especially important as the root-mean-square surface roughness 
angle exceeds the Cerenkov width, 
i.e., at high observing frequencies ($\nu \gtwig 300$ MHz, 
see section \ref{roughness}). 
On the other hand,
we show below that 
whenever there is significant contribution from ``upward'' neutrinos 
(neutrinos that
have survived a significant secant of lunar rock
and approach the surface from an upward
angle),
surface roughness plays a less important role.

Accounting for surface roughness may be accomplished 
by using a probabilistically smoothed $\scriptt_R$ function
defined by
\beq
\scriptt_R \ = \ \pi^{-3/2} \int_{-\pi/2}^{\pi/2} d\phi' \
\int_{-\infty}^{\infty} dw \ e^{-w^2} \scriptt_{w,\phi'}
{{({\hat n}_\sigma \cdot {\hat \gamma})} \over {({\hat n} \cdot
{\hat \gamma})}} \ \cong \  2 \pi^{-3/2} \int_{0}^{\pi/2} d\phi' \
\int_{-\infty}^{\infty} dw \ e^{-w^2} {\scriptt}_{w,\phi'}
 \ ,
\eeq
which allows the surface normal to be tilted, relative
to the radio ray incident along ${\hat \gamma}$, by a small
polar angle $\sigma = \sigma_o w$ in
a random azimuthal ($\phi'$) direction.
The polar angle $\sigma$ is approximately
normally distributed with Gaussian
halfwidth $\sigma_o$, as supported by the findings
of \citet{Shepard:1995} in the limit of small $\sigma$.
The expression 
$({\hat n}_\sigma \cdot {\hat \gamma}) / ({\hat n} \cdot {\hat \gamma})$
accounts for the higher likelihood of encountering a tilt that
{\it reduces} the angle of incidence than one that 
{\it increases} it, owing
to the relative increase in projected area of the former,
but this effect is deemed to be too negligible to track when the
tilt angles are small, consistent with our other approximations.
Here ${\scriptt}_{w,\phi'}$ expresses the no-internal-reflection
selection rule in the 
form of a step function whose argument is positive whenever
the values of $w$ and $\phi'$ allow the ray to cross the surface, so
$\scriptt_R$ itself becomes 
a smoothed function rather than a formal Heaviside
function.

It should be noted that
we choose the somewhat nonstandard convention of 
allowing $w$ to be either positive or negative, 
corresponding to surface tilts that either help or hinder ray escape,
and take the azimuth of
the surface normal to wrap only from $-\pi/2$ to $\pi/2$ (further applying 
the left/right 
symmetry to restrict to one azimuthal quadrant from $0$ to $\pi/2$),
rather than
restricting the tilt angle to positive values and wrapping around the full
$2 \pi$ in azimuth.  
We choose this convention because it indicates more explicitly when the
tilt is helping or hindering escape, an issue to which the Heaviside functions
are highly sensitive.
Also note that ${\scriptt}_R$ is normalized to unity in the
purely hypothetical situation where
${\scriptt}_{w,\phi'}$ is always unity, as desired.

Including surface roughness can increase the aperture significantly,
but it also adds two additional dimensions to the
integral over
fractional phase space, and it introduces a new angular scale, $\sigma_o$,
to the several other important angular scales that appear in the calculation.
It is natural to expect this complication to be worthwhile whenever
$\sigma_o$ is appreciable relative to
the other angular widths that affect the aperture, for example when
observing at high frequencies where the Cerenkov cone is thin.
The scaling laws we
derive bear out this expectation.

\subsection{Isotropic neutrino flux}

As the current interest is primarily
in cosmologically distributed sources, rather than targeted sources, we 
focus on isotropic neutrino flux distributions $I(E)$, 
and simplify the form of the aperture.
Then we may write  the detection rate $\scriptr$ as
\beq
\scriptr \ = \ \int dE \ I(E) A(E) \ ,
\eeq
where the energy-dependent isotropic aperture is given by
\beq
\label{isoa}
A(E) = \amax \int d \vmbar \int d \omegagbar
\int  d \omeganbar \ 
e^{-\tau_\nu} \ {\scriptt_R} \ {\scriptt_D} \ .
\eeq
Again note that each fractional phase space integral
is by definition normalized to unity, 
and $\amax$ specifies the {\it units} of $A(E)$,
but not the {\it scale} of $A(E)$.
The great
majority of the fractional phase space will 
generally not contribute to neutrino
detections, and thus
$e^{-\tau_\nu}$, ${\scriptt_R}$, and  ${\scriptt_D}$
exert stringent constraints on
$A(E)$ in ways we will now calculate.

\subsection{Spherical symmetry simplifications}

The fractional phase space in eq. (\ref{isoa}) includes 7 total
dimensions, or 9 including a surface-normal tilt distribution,
so would certainly represent a daunting undertaking to compute
in closed form.
However, we receive welcome assistance 
from the basic spherical symmetry of the Moon, in concert with an isotropic
incident neutrino flux.
Capitalizing on that symmetry actually allows us to eliminate three
of the dimensions of the phase space (two because the lunar
aperture operates the same way as seen by observers from all directions,
and one because the lunar aperture seen by those observers
has an
axial symmetry around the Moon center).
This reduces the phase space (even with surface roughness) to 6 dimensions,
which is certainly more
tractable, though still requiring extensive use of approximations
to achieve a closed-form result.


\subsection{Aperture calculation from the lunar
volume-centered perspective}

Several choices are possible for coordinatizing the phase space, each
with its various computational advantages and challenges.
We adopt a Moon-centered perspective that scans in spherical
coordinates over the volume of the Moon by accounting only
for the distance $h$ below the lunar surface 
along the radial direction, and we coordinatize the 
incident neutrino
direction in terms of the glancing angle 
$\alpha$ that the neutrinos make to the surface
(so $\alpha$ is 
the complement of the angle of incidence to the normal),
so chosen because it
tends to be a small quantity.
Our convention is $\alpha > 0$ 
for upward neutrinos that must penetrate a significant lunar secant
before reaching the detectable zone.

Hence, the 4-dimensional configuration phase space is divided into
one dimension to describe the location of interest in the Moon, 
one dimension to describe the incident neutrino angle,
and two dimensions to describe the outgoing electric field rays.
At this point in the calculation, we temporarily consider the two
dimensions of outgoing radio rays to be outside the Moon, but shortly
we will convert to a coordinatization where these are inside the Moon
and along the Cerenkov cone of the shower.
As mentioned above, in addition to these four dimensions, there are generally
two more integrations to account for the local surface normal variations
on the scale of the assumed
surface roughness.

The crucial simplification stemming 
from the spherical symmetry and the isotropic
neutrino assumption is that  fractional phase-space 
component weights are based only on the probability that a hypothetical
observer at Earth distance {\it randomly chosen} over the solid
angle of the lunar sky 
could detect the radio signal, without specifying any 
particular viewing angle.
This succeeds because we are in effect fixing the location where the
shower occurs along an arbitrary
radial ray in the Moon, randomizing the orientation
of the observer, and asking what is the probability 
that such a randomly located
observer could detect that shower.

Translating the above into an expression for the aperture yields
\beq
\label{ae2}
A(E) \ = \ \amax(E)
\times {1 \over 2} \int_{-\pi/2}^{\pi/2} d \alpha \ \cos \alpha \ 
\times\ \int d \omegagbar \ \times\ 
{3 \over R_m^3} \int_0^R dr \ r^2 \ e^{-\tau_\nu} \scriptt_D
\ \scriptt_R \ ,
\eeq
where $R_m$ is the radius of the Moon. 
Note how each fractional phase space component is still normalized to
unity, in keeping with the partition of $\amax$ over its contributing
fractional phase space.
Defining 
the neutrino interaction length
to be $\lnu(E) = 1/\rho \kappa(E)$ for cross
section per gram $\kappa(E)$ and mass density
$\rho$ (and we take $\rho = 1.8$ g cm$^{-3}$
in the regolith where the showers occur),
we obtain from eq. (\ref{emass})
\beq
\label{maxE}
\amax(E) \ =  4 \pi M\kappa(E) \ = 4\pi\left(
{4 \pi \over 3} R_m^3 \rho\right)\kappa(E)
\ = \ {4 \over {3}} \ {R_m \over \lnu}\asphere \ ,
\eeq
where 
\beq
\label{aodef}
\asphere \ = \ 4 \pi^2 {R_m}^2
\eeq 
is the maximum attainable aperture for a spherical
object for which neutrinos can be detected at most once (see
the Appendix for the reason that this ``maximum'' aperture could
actually be exceeded by added contributions from neutrinos with
initial energies above $E$).
Note that $\asphere$ is the geometric lunar cross section $\pi {R_m}^2$ times
the full 4$\pi$ steradians of illumination, as in \citet{Williams:2004},
and eq. (\ref{ae2})
yields $A(E) = \asphere$ if we set $\scriptt_D \scriptt_R = 1$ and
treat $\tau_{\nu}$ in the highly opaque limit, where
there is as yet no correction for
downgrading of higher-energy neutrinos (again,
see the Appendix for how the
effective $\tau_\nu$ is reduced by downgrading).

Since we treat only neutrino energies for which the Moon is highly opaque,
we have
$\asphere \ \<< \ \amax(E)$, owing to the high shielding of the
lunar interior,
and thus $\asphere$ makes for a better
fiducial reference to use in our aperture expressions. 
Thus we will express the aperture $A(E)$ in the form
\beq
\label{aperform}
A(E) \ = \ \asphere
\ P(E) \ ,
\eeq
where $P(E)$ is to be interpreted as the fraction of neutrinos 
entering the Moon at energy $E$ that will actually be detected
by the instrument under consideration,
assuming all such neutrinos will create showers, and corrected slightly
for the detection of downgraded neutrinos originally
at higher $E$ (see the Appendix).

\subsection{Converting from exterior to interior ray angles}

Since detectability of the radio rays is a crucial issue, the field
strengths generated by the hadronic showers appear prominently in
the aperture calculation.
However, the characteristics of these fields trace the Cerenkov cone
inside the lunar material, whereas the ray-angle phase space
is referenced to the emergent solid angle outside the Moon where the
detector is located.
This is inconvenient, so we now convert the ray angles to being inside
the Moon where the field properties are more easily expressed.
To effect that change, we must not only convert $d \omegagbar$ to
an angular integral over an {\it interior} solid angle, we must
also account for the solid-angle magnification factor that appears
whenever emerging rays refract across a dropping index of refraction,
which here falls from its internal value of $n_r \cong 1.73$ \citep{James:2009} to unity.
This magnification factor is \citep[e.g.,][]{Gorham:2004}
\beq
\label{xidef}
\xi(\beta) \ = \ {n_r^2 \cos \beta \over \sqrt{1 - n_r^2
\sin^2 \beta}} \ ,
\eeq
where $\beta$ is the angle of incidence from the normal of the radio ray
as it encounters the lunar surface from the {\it inside}.

We coordinatize this interior-ray solid angle using the polar
angle $\Delta$ (measured relative to the Cerenkov angle $\theta_c$ so
the $\Delta$ of interest are usually quite small), and the
azimuthal angle $\phi$ around the Cerenkov cone (with the convention
that $\phi=0$ corresponds to the direction nearest to the surface).
Hence we replace the normalized solid angle $d \omegagbar$ by
the transformed
solid angle $d\Delta \ \sin(\theta_c + \Delta) d\phi 
\ \xi(\beta) / 2 \pi$,
where we take $\phi$ only from 0 to $\pi$ because of the left/right
local symmetry.
The interior phase space element has its normalization altered by
the $\xi(\beta)$ factor, but the resulting exterior solid angle will not
exceed its full unit value because the contributing interior solid angle
is actually quite small, 
owing to truncation by $\scriptt_R$ as seen below.

Indeed we expect all integrals to be truncated
by the selection rules imposed by $\scriptt_R$ and $\scriptt_D$, so as a
minor convenience we
will extend all finite integration limits to infinity.
This step has no physical significance and will not alter the outcome of
the calculation, it merely removes unnecessary emphasis from the arbitrary
limits of the integrals, and returns the emphasis to the selection rules
themselves.
Since we are working in the limit where the radio rays are rapidly
attenuated in the Moon, we assume all detections occur near the surface,
so we 
also replace the integral over 
$r$ with an integral over $z = h/\lgamma$, 
where
$h$ is the depth below the surface and $\lgamma$ is the electric field
dissipation length
(so $\lgamma/2$ is the photon mean-free-path).
We also approximate $r^2$ by $R_m^2$
in the $z$ phase-space integration.
These are all excellent approximations in the domain of interest
of our calculation.

Taking the definition of $P(E)$ from eq. (\ref{aperform}), we
combine eqs. (\ref{ae2}) and (\ref{maxE})
to yield 
\beq
\label{probint}
P(E) \ = \ {1 \over \pi} \ {\lgamma \over \lnu} \
\int_{-\infty}^{\infty} d \alpha \ \cos \alpha \
\int_{-\infty}^{\infty} d \Delta \ \sin(\theta_c + \Delta) \
\int_0^\infty d \phi \
\int_0^\infty dz \ e^{-\tau_\nu} \ \scriptt_R \ \scriptt_D \ \xi \ .
\eeq
Note that already the scale of $A(E)$ has been
reduced by the factor $\lgamma / \lnu \ll 1$, the proportion by which
neutrinos overpenetrate to depths
beyond where detectable electric fields can emerge.
The remaining factors that will further
reduce the aperture, mediated by the $\scriptt$ truncations, will
be included next, and will invoke
additional approximations.

\subsection{Imposing the $\scriptt_R$ and $\scriptt_D$ selection rules}

The constraint described 
by $\scriptt_{w,\phi'}$
is that the radio ray emergent from the
electron shower must not internally reflect at the surface of the Moon,
and the constraint described by $\scriptt_D$ is
that the signal be detectable by the instrument of interest.
Each Heaviside function
uses whichever quantity must be positive to apply the constraint, which
appropriately truncates the
limits of integration.

We first consider the requirement that we count the contribution only
from rays that are associated with detectably
strong radio waves.
We assume that each shower induces a Cerenkov cone
with a field strength along each ray 
that depends on the deviation $\Delta$ in polar angle 
from the Cerenkov peak angle $\theta_c$.
If the field strength $\field$ is distributed over $\Delta$ in
an approximately Gaussian way \citep[small
deviations from this
are discussed by][]{Scholten:2006, Gusev:2006}, then
\beq
\label{fieldeq}
\field \ = \ \field_o  \Tpar(\beta) \ 
e^{-(\Delta/\Delta_o)^2} \ e^{-\tau_\gamma} \ ,
\eeq
where $\Delta_o$ is the Gaussian halfwidth of the angular distribution around
the Cerenkov angle, $\field_o$ is the strength of the field at the shower
along the Cerenkov angle,
$\tau_\gamma$ is the number of radio
dissipation lengths the field passes through before exiting the Moon at
the frequency in question, and $\Tpar(\beta)$ is the field transmission
coefficient
appropriate for the
diverging rays that fill the emergent solid angle
of interest.
We assume the
polarization is in the plane of incidence (termed ``pokey'' 
electric polarization), as that is the
dominant polarization for the rays most likely to escape total
internal reflection, and $\beta$ is the angle of incidence to the
surface normal (inside the Moon).

The result begins with the standard expression for the field
transmission coefficient for
plane waves \citep{Williams:2004}
\beq
\label{tpar}
\tpar \ = \ \sqrt{{n_r \cos \beta \over \cos \beta_o} \ \left (1 \ - \ \rpar^2 \right)}
\eeq
where $\rpar$ is the field reflection coefficient for ``pokey''
electric polarization,
and $\beta_o$ is the angle of refraction relative to the
normal (outside the Moon) 
as the rays pass through the surface of the Moon into free space,
so 
\beq
\beta_o \ = \ \sin^{-1}(n_r \sin \beta) \ .
\eeq
In an appendix of \cite{Williams:2004}, a 
ray-tracing technique is described
for converting this plane-wave
transmission coefficient to one appropriate for
the diverging rays that fill the observable emergent solid angle,
which agrees with an analytic result she quotes from a private
communication with Dave Seckel.
The result is
\beq
\label{Tpar}
\Tpar \ = \ \sqrt{{\tan \beta \over \tan \beta_o} \left (
1 \ - \ \rpar^2 \right )} \ = \ {\tan \beta \over \tan \beta_o}
\tpar(\beta) \ = \ {2 \sin \beta \cos \beta_o \over
\sin(\beta_+\beta_o)\cos(\beta-\beta_o)} \ ,
\eeq
where
\beq
\tpar(\beta) \ = \ {2 \sin \beta_o \cos \beta \over
\sin(\beta_o + \beta) \cos (\beta_o - \beta)} \ .
\eeq

It may be noted that the \cite{Williams:2004} numerical
result, the above analytic expression, and the analytic expression
cited in
\cite{Gusev:2006}, present sequentially more pessimistic
transmission coefficients, at roughly the 10\% level per step in the
sequence.
This motivates both our using the ``intermediate'' level of optimism,
and the application of the rather crude approximation outlined below.
It appears that the transmission of diverging radio rays from a hadronic shower
remains a problem that is not completely solved.

In this paper we use 
expressions for the maximum electric field and the Cerenkov cone 
width in the regolith as given by \cite{James:2009}, which tend to
give narrower detection windows and lower apertures than previous values
in the literature. 
For the maximum field we have
\beq
\label{eqn:e0}
\field_o(d,\nu,E) = 0.0845\ \frac{\rm{V}}{\rm{m}\cdot\rm{MHz}}
\ \left[\frac{d}{m}\right]^{-1}
\ \left[\frac{E_s}{EeV}\right]
\ \left[\frac{\nu}{\rm{GHz}}\right]
\ {\left[1+\left(\frac{\nu}{\rm{2.32~GHz}}\right)\right]}^{-1.23}\ ,
\eeq
where $d$ is the distance from the shower (and $m$ is the unit in
meters), 
$\nu$ is the observing frequency, and $E_s$ is the shower 
energy in EeV ($10^{18}$ eV). Approximately 20\% of the incident 
neutrino energy is deposited in hadronic showers, independent 
of neutrino flavor \citep{James:2009}, so we take $E_s = 0.2 E$.
The Cerenkov cone $1/e$ half-width is given by 
\beq
\Delta_o = 0.05 
\ 
\left[
\frac{\rm{GHz}}{\nu}
\right]
\ 
{\left[
1+0.075\ {\rm log}
\left(
\frac{E_s}{10^{19} \rm{eV}}
\right)
\right]}^{-1}\ .
\eeq

Note that we have multiplied the angular width 
constant $C_H = 2.4\degr$ of
\citep{James:2009} by the factor $1/\sqrt{ln \ 2} = 1.2$
to make $\Delta_o$ be
the $1/e$ half-width 
rather than the half-width at half-maximum, as this
is more attuned to use with the familiar exponential attenuation factors.
We also convert the angle units to radians for use in the scaling laws
that follow, since radians
are the useful unit for testing the small-angle approximations
that will be invoked shortly.

For a telescope with a field detection threshold
$\field_{min}$, the selection rule to have a detectable signal in
eq. (\ref{fieldeq}) is
\beq
\field_{min} \ < \ \field_o \Tpar(\beta)
e^{-(\Delta/\Delta_o)^2} \ e^{-\tau_\gamma} \ ,
\eeq
or, solving for $\Delta$,
\beq
\label{detect}
\Delta \ < \ \Delta_o \sqrt{{\rm ln} \left ( {\Tpar(\beta)
\field_o \over \field_{min}} 
\right ) \ - \ {s \over \lgamma} }\ ,
\eeq
where $s$ is the path length from the shower to the surface and $\lgamma$
is the electric field dissipation length.
Therefore ${\scriptt}_D = 0$ when this is not satisfied, 
which truncates
the integrals over $\Delta$ and
$z$ in a manner that depends on the order of integration.
When the above expression is satisfied, 
the ray in question is detectable,
and ${\scriptt}_D = 1$.

To avoid internal reflection, the ray must meet the surface at an angle of
incidence that exceeds
the complement of the Cerenkov angle, where the Cerenkov
angle is $\theta_c = \cos^{-1}(1/n_r) = 0.954$ for $n_r = 1.73$.
The transmission coefficient $\Tpar(\beta)$ goes to zero at that
angle, which
is when $\beta = \pi/2 - \theta_c$, and remains zero for all larger 
$\beta$.
Thus, the constraint in eq. (\ref{detect}) is already violated when
internal reflection occurs, and there would be
no formal need
to include $\scriptt_{w,\phi'}$ separately.
However, $\Tpar$ grows so rapidly with angle 
that it is appreciably nonzero even just
a few degrees from critical, and it quickly
saturates near 0.7 for angles as small as 10 degrees from critical.
That rapidly saturating behavior, along with the fact that $\Tpar$
appears in a logarithm so is only crucially important for weaker fields,
motivates our choosing to model $\Tpar \cong 0.6$ as a constant.
This loses some accuracy in the result, but is much computationally
simpler than following a function whose small-angle approximation breaks
down completely for angles more than just 2 degrees from critical.
Uniformly setting $\Tpar = 0.6$ 
implies that we no longer have zero transmission
at the critical angle, so we now need to apply the total internal
reflection constraint, $\scriptt_{w,\phi'}$, separately.

To determine if internal reflection is avoided, we need the cosine of 
the angle
of incidence, which is
given by the dot product of the surface
normal $\hat{n}(w,\phi')$ and the 
ray direction $\hat{\gamma}(\Delta,\phi,\alpha)$.
Here $\alpha$ is the angle the neutrino (and its shower) makes to
the horizontal, where our convention is that $\alpha < 0$ is for downward
neutrinos that are first encountering the Moon close to the shower of 
interest.
Thus given the glancing incident neutrino angle $\alpha$, which sets
the orientation of the Cerenkov cone, and the 
electric field ray
angles $\Delta$ and $\phi$ within that
cone (where $\Delta = 0$ is at the Cerenkov angle
and $\phi=0$ is most directly toward the surface
and least apt to internally reflect), we require
\beq
\hat{n}(w,\phi') \cdot \hat{\gamma}(\Delta,\phi,\alpha) 
\ > \ \sin \theta_c \ ,
\eeq
which yields directly
\begin{multline}
\label{reflect} 
\sin \theta_c \ <  \sin(\theta_c+\Delta)\cos \phi \sin \sigma
\cos \phi' \sin \alpha \ - \ \cos(\theta_c+\Delta)\cos \sigma \sin \alpha
\\
\ - \ \sin(\theta_c+\Delta) \sin \phi \sin \sigma \sin \phi' \ + \
\sin(\theta_c+\Delta) \cos \phi \cos \sigma \cos \alpha 
\\
\ + \
\cos(\theta_c+\Delta) \sin \sigma \cos \phi' \cos \alpha  \ .
\end{multline}
Applied to the integrals in eq. (\ref{probint}),
this constraint identifies a maximum azimuth $\phi$ 
and truncates the $\phi$ integral there, while
the constraint in eq. (\ref{detect})
truncates the integral over depth $z$.

\section{Further approximations to reach a result in closed form}

The goal of this paper is to achieve closed-form 
expressions for the effective area $A(E)$ 
as a function of neutrino energy, lunar
regolith properties, and observing parameters.  
Since at this point we still have a cumbersome 6-dimensional 
integral to evaluate, we must avail ourselves of several additional
approximations to obtain a tractable expression which yields 
the desired scaling laws.

\subsection{The near-surface emission approximation}

Since the integrand $e^{-\tau_\nu}$ depends on $z$,
the $z$ integration is nontrivial,
but this dependence is removed when the observable
showers have to be so close to the surface that there
is no appreciable change in the neutrino irradiation
over the region where the radio signals are detectable.
The requirement to make this approximation is 
\beq
\lgamma \ \<< \ {\lnu^2 \over R_m} 
\eeq
where $\lgamma$ and $\lnu$ are the 
electric-field attentuation length 
and neutrino mean-free-path respectively.
This ``near-surface emission'' approximation allows the
$z$-integral to be performed trivially.
The expressions for $\lnu$ and $\lgamma$ are taken as
\citep[summing the cross sections for neutral and charged-current interactions, as they both generate similar hadronic showers,][]{Gandhi:1998,Reno:2005},
\beq
{\lnu \over R_m} \ = \ 0.07 \left ( {E \over 10^{20} eV}
\right )^{-1/3} \left ( \rho \over {\rho_{reg}} \right )^{-1} \ ,
\eeq
where $\rho_{reg} = 1.8$ g cm$^{-3}$ is the regolith mass density, and
\beq
{\lgamma \over R_m} \ = \ 
1 \times 10^{-5} \left ({\nu \over GHz}\right )^{-1} \ .
\eeq
(This result is known only to within about a factor of two \citep{Gorham:2004,Scholten:2006}.
Hence the approximation requires $\nu \  \>> \ 5
 (E/10^{20}$ eV$)^{2/3}$ MHz for lunar rock with density $\rho \cong
3$ g cm$^{-3}$.
This is
satisfied over most of the range of neutrino energies and radio
frequencies that concern us, 
except when the lowest frequencies ($\nu \ltwig 100$ MHz) are used to
observe the highest energy ($E \>> 10^{22}$ eV) neutrinos.

The surface-emission approximation allows us to replace $\tau_\nu$
in $e^{-\tau_\nu}$ by
\beq
\label{tauexpression}
\tau_\nu \ \cong \ {R_m \over \psi \lnu} 
\sin \alpha \ \scriptt(\alpha)
\eeq
for $\scriptt(\alpha)$ again the Heaviside step function, so $\tau_{\nu} = 0$
for $\alpha < 0$ (downward neutrinos) and 
$\tau_{\nu} \ = \ 
\sin \alpha \times 2R_m/\lnu$ for $\alpha > 0$ (upward neutrinos).
Here we insert the order-unity parameter $\psi = 1.4$ (appropriate
for a neutrino spectrum $\sim E^{-2}$, see Appendix)
to account for neutrinos from higher energies being downgraded into
the detection regime for energy $E$, which slightly enhances the aperture.
Note the absence of dependence on $z$, making the $z$ integration
a trivial matter of simply tracking its range of contribution.

\begin{figure}[h]
\begin{center}
\includegraphics[width = 6in]{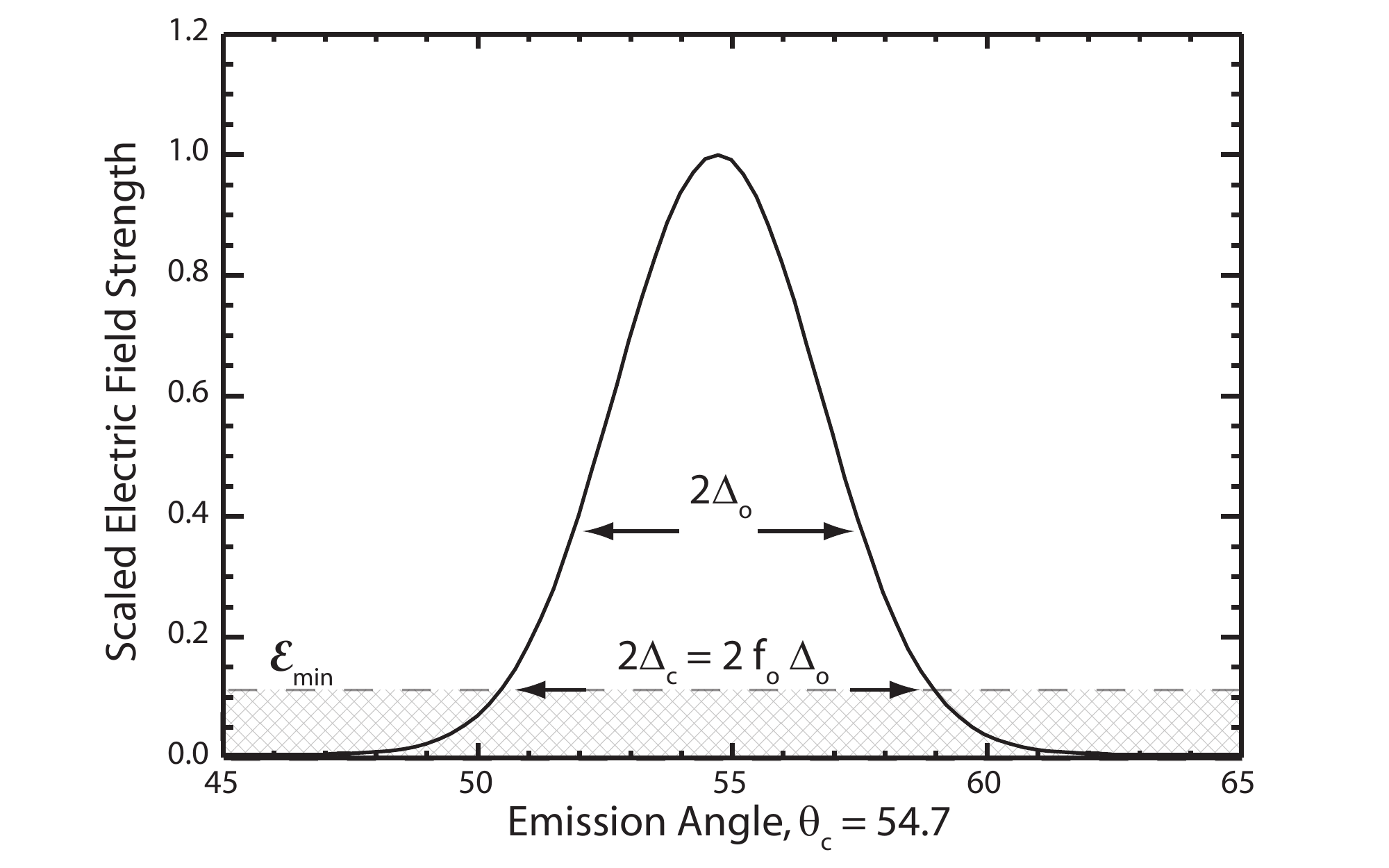}
\caption{Cerenkov cone angular profile at $E= 10^{21}$ eV and $\nu =$ 1~GHz. The dimensionless factor $f_o$ is the ratio of the cone thickness at threshold $\field_{min}$ to the $1/e$ width.\label{Fig:Conewidth}}
\end{center}
\end{figure}

Carrying out the $z$ integration subject to the selection rule
in equation (\ref{detect}) and using  
\beq
\label{smax}
{s \over \lgamma} \ = \
{z \over \sin(\theta_c+\Delta)\cos \phi \cos \alpha \ - \ 
\cos(\theta_c+\Delta)\sin \alpha} \ ,
\eeq
the detection probability integral (eq. [\ref{probint}]) can be written
\begin{multline}
\label{afterz}
P(E) \ = \ {2 \over \pi^{5/2}} \ {\lgamma \over \lnu} \
\int_{-\infty}^{\infty} d \alpha \ \cos \alpha \
\int_{-\infty}^{\infty}
d \Delta \ \sin(\theta_c + \Delta) \
 \\
\times \int_0^\infty d \phi \ \int_0^{\pi/2} d\phi' \ \int_{-\infty}^{\infty}
dw e^{-w^2}
\Upsilon(z_{max}) \ e^{-\tau_\nu} \
\scriptt_{w,\phi'} \ \xi(\beta) \ ,
\end{multline}
where $z_{max}$ is the maximum depth for which the emergent field 
exceeds the minimum detectable value, and we define
\beq
\Upsilon(f) \ = \ f \scriptt(f)
\eeq
for $\scriptt$ the Heaviside step function,
to insure the integrand never contributes when it is
negative.
We may determine $z_{max}$
by combining eqs. (\ref{detect}) and (\ref{smax})
\beq
z_{max} \ = \ 
[
\sin(\theta_c + \Delta)\cos \phi \cos \alpha \ - \ \cos(\theta_c+\Delta)
\sin \alpha ]
\ f_o^2 \left ( 1 \ - \ {\Delta^2 \over f_o^2
\Delta_o^2} \right )
\eeq
where $f_o$ is the dimensionless quantity
\beq
f_o \ = \ \sqrt{{\rm ln} \left ( {\field_o \Tpar \over \field_{min}} \right ) } \ .
\label{fodef}
\eeq
Physically, $f_o$ is the ratio of the thickness of the Cerenkov 
cone corresponding to the electric field threshold to the $1/e$ full 
thickness ($2\Delta_o$), as shown in Fig.~\ref{Fig:Conewidth}. 

\subsection{Small angle approximation}

We have now reduced the number of integrations to five, including two
over surface roughness.
One more integration, that over azimuthal angle
around the Cerenkov cone
$\phi$, can be carried out surprisingly easily
if we make
the further approximation that the detectable radio rays hug
tightly to the Cerenkov cone, such that $\Delta$ is small, and
that the other angles we treat,
$\phi$, $\alpha$, and $\sigma$, are also small.
The small-angle approximation is 
particularly valid  for lower energy neutrinos
near the important GZK cutoff.
It also applies to higher observing frequencies above 1 GHz,
and may extend to lower frequencies as long as the highest
energy neutrinos are not the focus.
In this approximation, the maximum azimuthal angle that does not
totally internally reflect, $\phi_R$, is
given by equation (\ref{reflect}) to be
\beq
\label{phir}
\phi_R \ = \ 
\sqrt{{2 \left(\Delta \ + \ w \sigma_o \cos \phi' 
\ - \ \alpha\right) \over \tan \theta_c}} \ \ ,
\eeq
which supplies the upper limit for the $\phi$ integration.

The small-angle approximation also permits us to use
\beq
\label{zmaxeq}
z_{max} \ \cong \ \sin \theta_c \ f_o^2 \left ( 1 \ - \ 
{\Delta^2 \over f_o^2 \Delta_o^2} \right ) \ ,
\eeq
and to replace $\sin \theta$ by $\sin \theta_c \ = \ \sqrt{1 - 1/n_r^2}$ 
and $\cos\alpha$
by unity.
This gives us all the expressions we need to simplify the evaluation
of the integrals, recalling again that $f_o$ is influenced by $\Tpar$,
and to within an expected accuracy of
10-20\% we
take $\Tpar(\beta) = 0.6$, 
rather than its small-angle form, as the
latter loses accuracy too rapidly for the angles we need to treat.
We note that the primary influence of $\Tpar$ is in determining the 
depth below the surface that will be visible, 
and since the angular contributions scale in
a self-similar way as this depth is varied at different $E$,
the systematic errors introduced by our approach should appear primarily
in the energy scale.
Hence the transmission coefficient has its largest significance in the
determination of the minimum detectable neutrino energy,
and since this is already an important issue for GZK-type neutrinos, future
work should attempt to clarify the reliability of the various
and somewhat contradictory treatments
of $\Tpar$ found in the literature.

\subsection{Evaluation of the azimuthal integral over the Cerenkov cone}

Approximating all remaining angles to lowest nonvanishing order
simplifies the expressions dramatically, and allows us to carry out
the $\phi$ and $w$ integrations in closed form.
Prior to those integrations, and subject to the approximations above,
the detection probability becomes
\begin{multline}
\label{apersmall}
P(E) \ = 
{2 \over \pi^{5/2}} {(n_r^2 - 1) \over n_r^2}  f_o^2
\left( {\lgamma \over \lnu}\right)
\int_{-\infty}^\infty dw \ e^{-w^2} 
\int_0^{\pi/2} d\phi' \
\\ \times
\int_{-\infty}^{\infty} d \Delta \ 
\Upsilon\left ( 1 \ - \
{\Delta^2 \over f_o^2 \Delta_o^2} \right )
\int_{-\infty}^{\infty} d \alpha \ e^{-\tau_\nu}
\int_0^{\phi_R} d\phi \ \xi(\beta) \scriptt_R(\phi_R) \ ,
\end{multline}
where 
in the surface-emission approximation we have 
\beq
\tau_\nu \ = \ \Upsilon \left ({\alpha \over \alpha_o} \right )
\eeq
and
we have defined
\beq
\label{alphaodef}
\alpha_o \ = \ {\lnu \psi \over 2 R_m} \ = \ 0.03 \left (
{E \over 10^{20} eV} \right )^{-1/3}
\eeq
as the angle through which upward ($\alpha > 0$) neutrinos are successful at
penetrating the lunar secant through rock with mass density $\rho =$ 3 g cm$^{-3}$ up to the regolith layer (with $\psi \cong 1.4$
chosen in regard to a somewhat arbitrary neutrino spectrum with power-law
index -2, 
see Appendix).

To evaluate the $\phi$ integral,
let us first let us define the angle $\varepsilon$, which is 
the angle of incidence
(from inside the Moon) to the normal to the surface, relative to the
critical angle, so 
\beq
\varepsilon  \ = \ {\pi \over 2} \ - \ \theta_c \ - \ \beta \ .
\eeq
We replace $\beta$ by $\varepsilon$ because the latter is small and will
be considered only to lowest nonvanishing order.
To this order, we find eq. (\ref{xidef}) becomes
\beq
\label{xismall}
\xi(\varepsilon) \ \cong \ {n_r (n_r^2 - 1)^{1/4} \over \sqrt{2 \varepsilon}}
\ ,
\eeq
which demonstrates explicitly how the solid-angle magnification factor 
gets large as the critical angle is approached.
This magnification effect is an important contributor to what would otherwise
be a much smaller aperture.

The angle $\varepsilon$ is not one of the phase space integration variables, 
but is easily expressed in terms of those variables, in the
small-angle limit:
\beq
\label{epssmall}
\varepsilon \ \cong \ \Delta \ + \ w \sigma_o \cos \phi' \ - \
\alpha \ - \ {\sqrt{n_r^2 - 1} \over 2} \phi^2 \ = \ 
{\sqrt{n_r^2 - 1} \over 2} \phi_R^2 \left (1 \ - \ {\phi^2 \over \phi_R^2}
\right ) \ ,
\eeq
where in this limit we have
\beq
\phi_R \ = \ {2 (\Delta  +  w \sigma_o \cos \phi'  -  \alpha)
\over \sqrt{n_r^2 - 1}} \ .
\eeq
Now we encounter an interesting result, which follows from elementary
application of eqs. (\ref{xismall}) and (\ref{epssmall}):
\beq
\int_0^{\phi_R} d\phi \ \xi(\varepsilon) \scriptt(\phi_R) \ = \ 
{n_r \pi \over 2} 
\scriptt(\phi_R) 
\eeq
for any set of values for the other phase-space variables.
In other words, in each equal-size
phase-space bin over $\Delta$, $\alpha$, $\sigma$,
and $\phi'$, the integral over $\phi$ yields either 
{\it the same} fixed numerical
value, or it vanishes.
Physically, this says that the solid-angle magnification effect
perfectly compensates
for the way internal reflection truncates the azimuthal window for ray
escape.
This can be no coincidence, and presumably would be more intuitively
clear using some other choice of
phase-space partition.

\subsection{Evaluation of the integral over incident neutrino angle}

The simple result from the $\phi$ integration is especially helpful in
carrying out the integral over $\alpha$, because it adds no new dependence
on $\alpha$ to the integrand.
Thus the integrand remains simply
$e^{-\Upsilon(\alpha/\alpha_o)}$, which is amenable to closed-form
integration.
When $\alpha > 0$ (upward neutrinos), we have an exponential integral,
and when $\alpha < 0$ (downward neutrinos), we have a trivial integrand.

At this point it is convenient to introduce the scaled variables
\beq
v \ = \ {\alpha \over \alpha_o}
\eeq
and 
\beq
u \ = \ {\Delta \over f_o \Delta_o} \ ,
\eeq
and define the ``roughness parameter''
\beq
x \ = \ {\sigma_o \over f_o \Delta_o} \ ,
\eeq
which characterizes how the window of acceptance of downward
neutrinos can be expanded by surface tilts that avoid 
total internal reflection,
and the ``penetration parameter''
\beq
y \ = \  {\alpha_o \over f_o \Delta_o} \ ,
\eeq
which characterizes the contribution of upward neutrinos that penetrate
a long distance through the lunar rock to reach the shower point.
Using these definitions, 
eq. (\ref{apersmall}) becomes

\beq
P(E) \ = \ P_o(E) \alpha_o \int_0^{\pi/2} d\phi' \ 
\int_{-\infty}^\infty dw \ e^{-w^2} \int_{-1}^1 du \ (1-u^2)
\int_{-\infty}^\infty dv \ e^{-\Upsilon(v)}
\scriptt (v  -  v_{min} )
\eeq
where for brevity we define
\beq
P_o(E) \ = \  {(n_r^2 - 1) \over \pi^{3/2} n_r}  
\left( {\lgamma \over \lnu}\right) f_o^3 \Delta_o 
\eeq
and
\beq
v_{min} \ = \ -{u\over y} \ - \ {w x \cos \phi' \over y} \ .
\eeq
The integration over $v$ may be carried out explicitly, yielding
\begin{multline}
\label{paccurate}
P(E)\ = \ P_o(E)\ \alpha_o \int_0^{\pi/2} d\phi' \
\\ \times
\left [ \int_{w_o}^\infty \ e^{-w^2} 
\int_{-1}^1 du \ (1  -  u^2) U_{+} \ + \ 
\int_{-w_o}^{w_o} dw \ e^{-w^2} \int_{w/w_o}^1 du \ (1-u^2) U_{-} \right ]
\end{multline}
where
$w_o \ = \ 1/(x \cos \phi')$ and
\beq
U_{\pm} \ = \ 1 \ + \ {u \over y} \ \pm \ {w x \cos \phi' \over y}
\ + \ e^{\pm (u/y \ - \ w x \cos \phi'/w_o)} \ .
\eeq

Although further progress can be made carrying out the 
elementary integrals
over the Cerenkov cone angle $u$, the expressions become extremely unwieldy,
and the two further integrals over the surface tilt 
parameters $w$ and $\phi'$ would require numerical integration anyway.
So we pursue no further the formal integration of these
expressions, and instead
turn to calculating scaling laws in various asymptotic limits.
We ultimately find that a convenient global approximation 
may be derived without significant loss of accuracy,
by using the approximate method described next.

\subsection{Asymptotics and scaling laws}

Since the ultimate goal is to gain insight by deriving scaling laws
and asymptotic limits, we seek an approximate evaluation of
eq. (\ref{paccurate}) that is accurate in both
limits of weak and strong roughness ($x\rightarrow0$ and $x \ \>> \ 1$,
respectively), which will then hopefully maintain approximate accuracy
in between.
We may accomplish this simply by evaluating the aperture in these
two opposite extremes, keeping only the leading contributions in each regime,
and simply adding them together for a global scaling law.
Because the regimes of contribution are so physically separable, simply
adding them produces results that agree with the full expressions within
$\sim$25\%, and also yields physically insightful results.

Taking $x=0$ produces closed-form results for the integral
in eq. (\ref{paccurate}), but even those results are more complicated
than necessary.
Keeping only the dominant terms in the limit of either large or small
$y$, we find we achieve better than 25\% accuracy with the 
simple asymptotic expression
\beq
P(E)_{smooth} \ \cong \ {(n_r^2 - 1) \over 8 n_r} 
{\lgamma \over \lnu} f_o^3 \Delta_o \left (
f_o \Delta_o \ + \ {16 \over 3} \alpha_o 
\right ) \ .
\eeq
Note the first term in the parentheses originates from downward neutrinos
and the second from upward neutrinos, so we see the angular acceptance of
downward neutrinos is controlled by the Cerenkov width when the surface
is smooth, and the acceptance of upward neutrinos is due to the angle that
can successfully penetrate the Moon.

If we include roughness and consider the limit $x \ \>> \ 1$, we find
that the upward neutrino detection rate is hardly affected, because upward
detections are limited far moreso by penetration than 
by total internal reflection, giving them a completely different
character from downward detections.
However, the downward detection rate may be greatly enhanced by
roughness, because downward
neutrinos often produce Cerenkov cones that are almost 
completely internally
reflected unless the surface encountered by the rays is 
favorably tilted, and we find in
this limit
\beq
P(E)_{rough} \ \cong \  {(n_r^2 - 1) \over 8 n_r}
{\lgamma \over \lnu} f_o^3 \Delta_o 
\left(
{16 \over 3 \pi^{3/2}}\sigma_o 
\right)
\ ,
\eeq
so the role of $f_o \Delta_o$ for the downward neutrinos has been supplanted
by $\sigma_o$; rays avoid internal reflection not by pushing to the edge of
the detectable Cerenkov cone, but rather by being lucky enough to encounter
the surface where the tilt is highly favorable.

Adding the smooth and rough limits thus yields the general expression,
consistent with pre-existing inaccuracies owing to the idealizations
and approximations already in place.
Thus we find for the full aperture calculation
\beq
\label{aperapprox}
A(E) \ \cong \ A_o  {(n_r^2 - 1) \over 8 n_r}
{\lgamma \over \lnu} f_o^3 \Delta_o
\left (\Psi_{ds} \ + \ \Psi_{dr} \ + \ \Psi_u \right ) \ ,
\eeq
where
\beq
\Psi_{ds} \ = \ f_o \Delta_o 
\eeq
accounts for downward detections without help from roughness,
\beq
\Psi_{dr} \ = \  {16 \over 3 \pi^{3/2}} \sigma_o \ = \   0.96 \ \sigma_o
\eeq
accounts for downward detections assisted by roughness,
and
\beq
\Psi_{u} \ = \  {16 \over 3} \alpha_o \ = \ 5.3 \ \alpha_o
\eeq
accounts for the detection of upward neutrinos.

The above results underline clearly 
the three basic angular scales of interest, the Cerenkov width
$\Delta_c = f_o \Delta_o$, the surface roughness parameter $\sigma_o$, and the
upward neutrino acceptance angle $\alpha_o$, which with
appropriate coefficients, map 
into the angular acceptance parameters $\Psi_{ds}$, $\Psi_{dr}$, and $\Psi_u$.
The aperture for neutrino detection is simply dominated
by whichever of these is largest, and note that the $\alpha_o$ angle
for upward neutrinos is benefited by an order-unity
coefficient that is significantly larger than the others, owing to the
much higher azimuthal angular acceptance when internal reflection is
less of a problem.
Nevertheless, the large scale of $\sigma_o$  ($>10\degr$ at $\nu\gtrsim1$ GHz)
makes roughness an important contributor, especially at higher frequencies,
whereas the $f_o \Delta_o$ parameter is largest at low frequencies.
Upward neutrinos, on the other hand, prefer lower energy neutrinos that
penetrate over a wider $\alpha_o$, but are often difficult to detect unless
the instrument is extremely sensitive.

\section{Aperture Dependence on Lunar and Telescope Parameters}

\begin{figure}[t]
\begin{center}
\includegraphics[width = 6in]{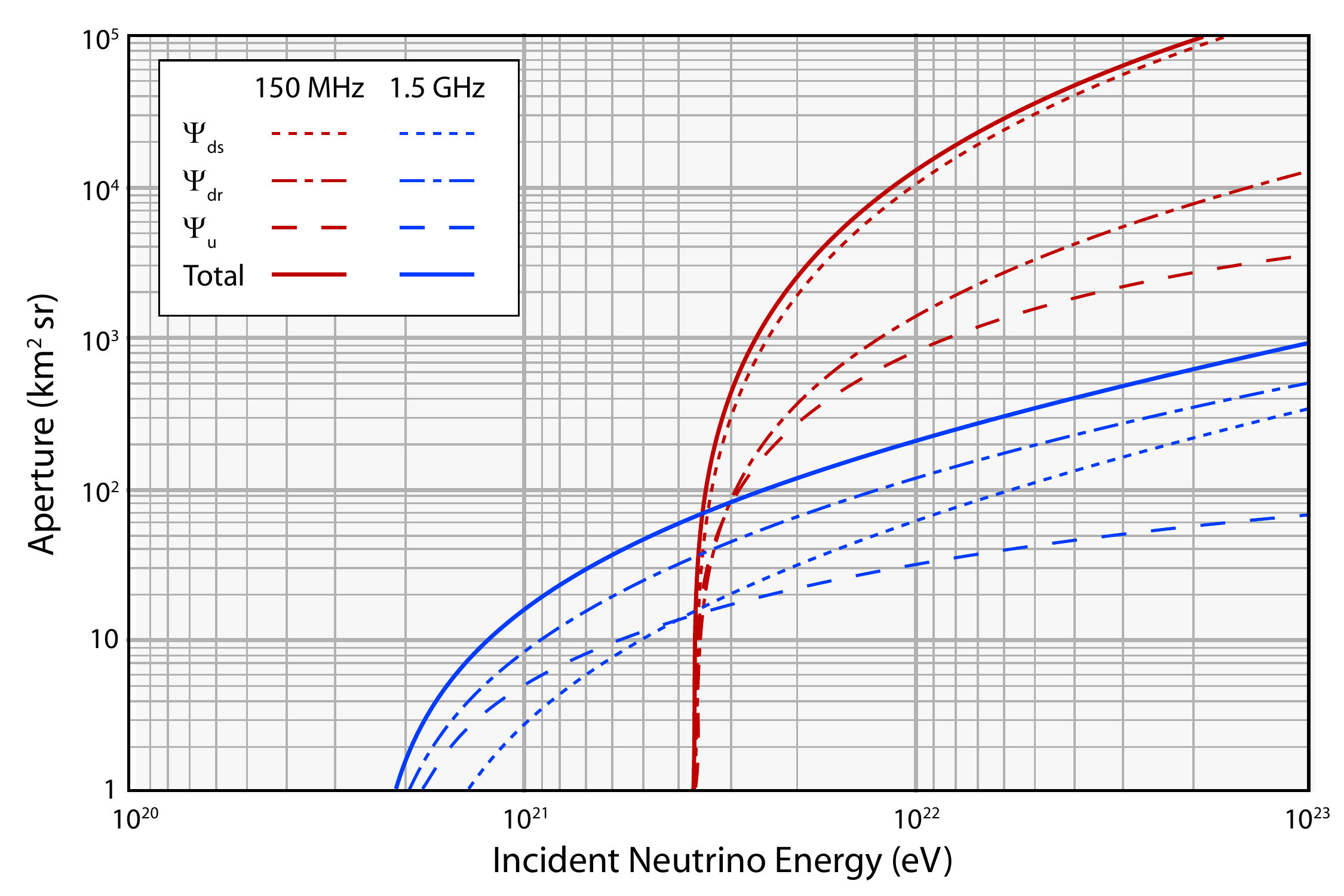}
\caption{Total aperture versus neutrino energy at threshold field $\field_{min}$ =\ 0.01\ \edim and two observing frequencies: 150~MHz (solid red line) and 1.5~GHz (solid blue line). The contribution from each of the three terms in equation~(\ref{aperapprox}) is also shown. At both frequencies,  downward-directed terms have the largest contribution except near the lower cutoff energy. For energies well above the cutoff energy, surface roughness has the largest contribution,  except at low frequencies, where the smooth surface term already
dominates without assistance from roughness. \label{Fig:Psi}}
\end{center}
\end{figure}

The scaling laws provided by
equation (\ref{aperapprox}) may be viewed as the fundamental
result of this paper.
Each of the three terms has a simple physical interpretation
that can be conceptualized in the following
simple form:
\beq
P_i \ = \ P_o \  \Psi_i 
\eeq
for $i = {ds, dr, u}$,
and where 
\beq
P_o \ = \ 
 {(n_r^2 - 1) \over 8 n_r}
{\lgamma \over \lnu} f_o^3 \Delta_o
\eeq
accounts for the inherent angular width of the Cerenkov emission
as well as the fractional loss in detections owing to the
overpenetration of neutrinos past the layers where radio rays may
be detected.
The $\Psi_i$ terms represent effective
acceptance angles for the ``active'' neutrinos, i.e.,
the neutrinos that can contribute to detections.
The relative contribution of these terms to the 
total aperture is illustrated in Fig.~\ref{Fig:Psi} 
for a threshold field $\field_{min} = 0.01$\edim and two 
characteristic frequencies, 150~MHz and 1.5~GHz. 

\subsection{Dependence on Surface roughness}
\label{roughness}

It is apparent from Fig.~\ref{Fig:Psi} that the surface 
roughness contribution $\Psi_{dr}$ is the largest contributor 
to the total aperture at  1.5~GHz,  whereas it is unimportant at 150~MHz.  
The roughness contribution can be very large at even higher frequencies e.g., 
the GLUE experiment at 2.2~GHz \citep{Gorham:2004}, 
where the enhancement over a smooth lunar surface is a more than 
a factor of 3 at $E = 10^{21}$ eV.  
The contribution from surface roughness is important when it exceeds 
the sum of the contributions from 
the smooth surface and upward-going contributions 
For $E_{\nu} = 1$ ZeV, this occurs at $\nu\gtwig$\ 300~MHz 
with only a modest dependence on neutrino energy. 
The importance of surface roughness to the detection aperture at 
high frequencies has also been found in several Monte Carlo 
ray-tracing simulations of the escaping radiation from the lunar 
surface \citep{Gorham:2001, Beresnyak:2003, James:2009}. 

\subsection{Dependence on minimum detectable electric field ($\field_{min}$)}
\label{min-efield}

\begin{figure}[t]
\begin{center}
\includegraphics[width = 4.5in]{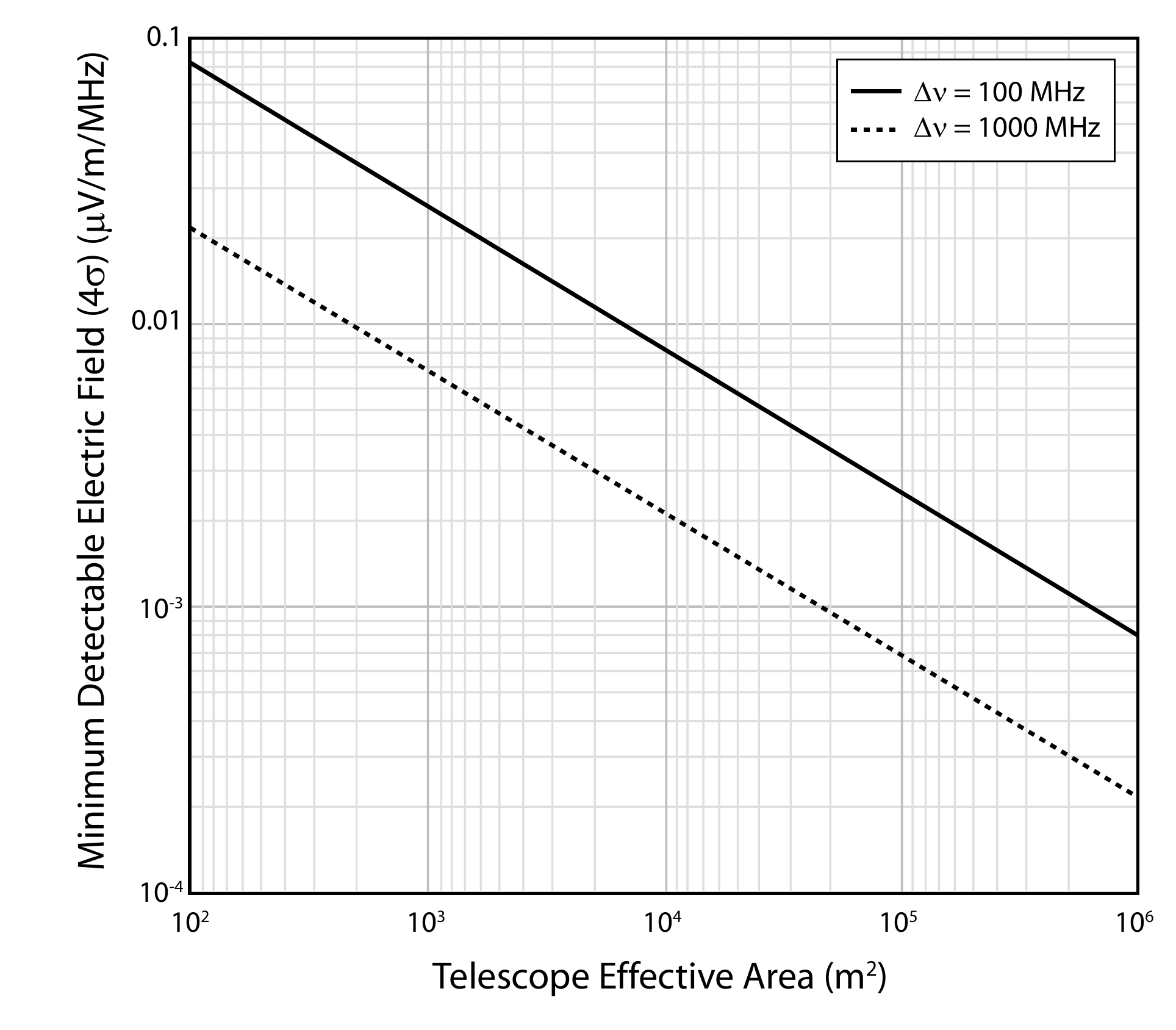}
\caption{Minimum detectable electric field vs telescope effective area 
using eq. (\ref{emin}) and bandwidth $\Delta\nu = $100~MHz (solid line) 
and 1~GHz (dashed line). 
See text for assumed values for other 
telescope parameters.\label{Fig:emin-plot}}
\end{center}
\end{figure}

The minimum detectable electric field of a radio telescope 
with effective collecting area $A_{e}$, system temperature $T_{sys}$, 
and bandwidth $\Delta\nu$, receiving linearly-polarized radiation, 
can be written \citep{Gorham:2004}
\beq
\label{emin}
\field_{min} = N_{\sigma}\
{
\left(
\frac
{2k_bT_{sys}Z_0}
{A_{e}\Delta \nu\ n_r}
\right)
}^{\onehalf}\ \ \rm{V/m}
\eeq 
where $N_{\sigma}$ is the minimum number of standard deviations 
needed to reject statistical noise pulses, $k_b$ is Boltzmann's constant, 
$Z_0$ = 377 $\Omega$ is the impedance of free space, 
and $n_r$ is the refractive index of the medium. 
This minimum electric field is plotted as a function effective 
telescope collecting area in Fig.~\ref{Fig:emin-plot}, 
using nominal values of $N_{\sigma}$ = 4.0, $T_{sys} = 120\degr$ K 
(assumed dominated by the Moon's contribution, limb pointing), 
and $n_r = 1.73$. 

\begin{figure}[h]
\begin{center}
\includegraphics[width = 6in]{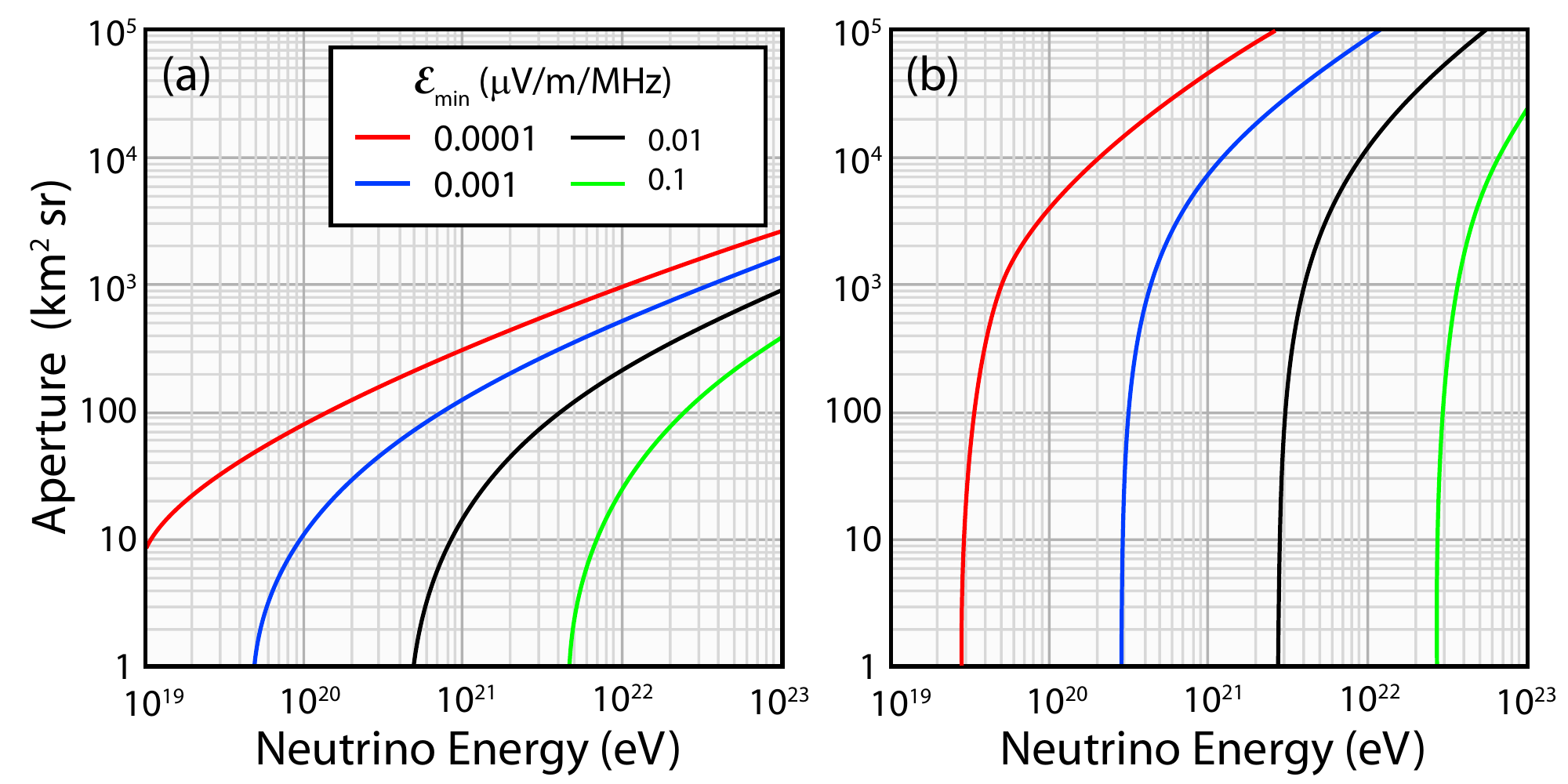}
\caption{$(a)$ Aperture vs. neutrino energy at observing 
frequency $\nu =$ 1.5~GHz using telescopes with minimum detectable 
electric field $\field_{min}$= 0.0001, 0.001, 0.01, and 0.1 \edim. 
$(b)$ Same as panel $(a)$, but at observing frequency 
$\nu= 150$~MHz. 
Note the trade-off between collecting area and minimum detectable 
neutrino energy as frequency is changed.\label{Fig:Emin-dF}}
\end{center}
\end{figure}

Since the effective thickness of the Cerenkov cone $\Delta_c$ depends 
strongly on $\field_{min}$, the total effective aperture will also 
depend strongly on $\field_{min}$, as shown in Fig.~\ref{Fig:Emin-dF}. 
The left panel shows the aperture at an observing frequency 1.5~GHz, 
while the right panel is at 150~MHz. 
This plot shows the clear 
trade-off between collecting area and minimum detectable neutrino 
energy as the observed frequency changes: 
at high frequencies the minimum neutrino energy cutoff is lower, 
while at low frequencies the total aperture is significantly higher 
at a given fixed telescope sensitivity ($\field_{min}$). 

\begin{figure}[t]
\begin{center}
\includegraphics[width = 6.5in]{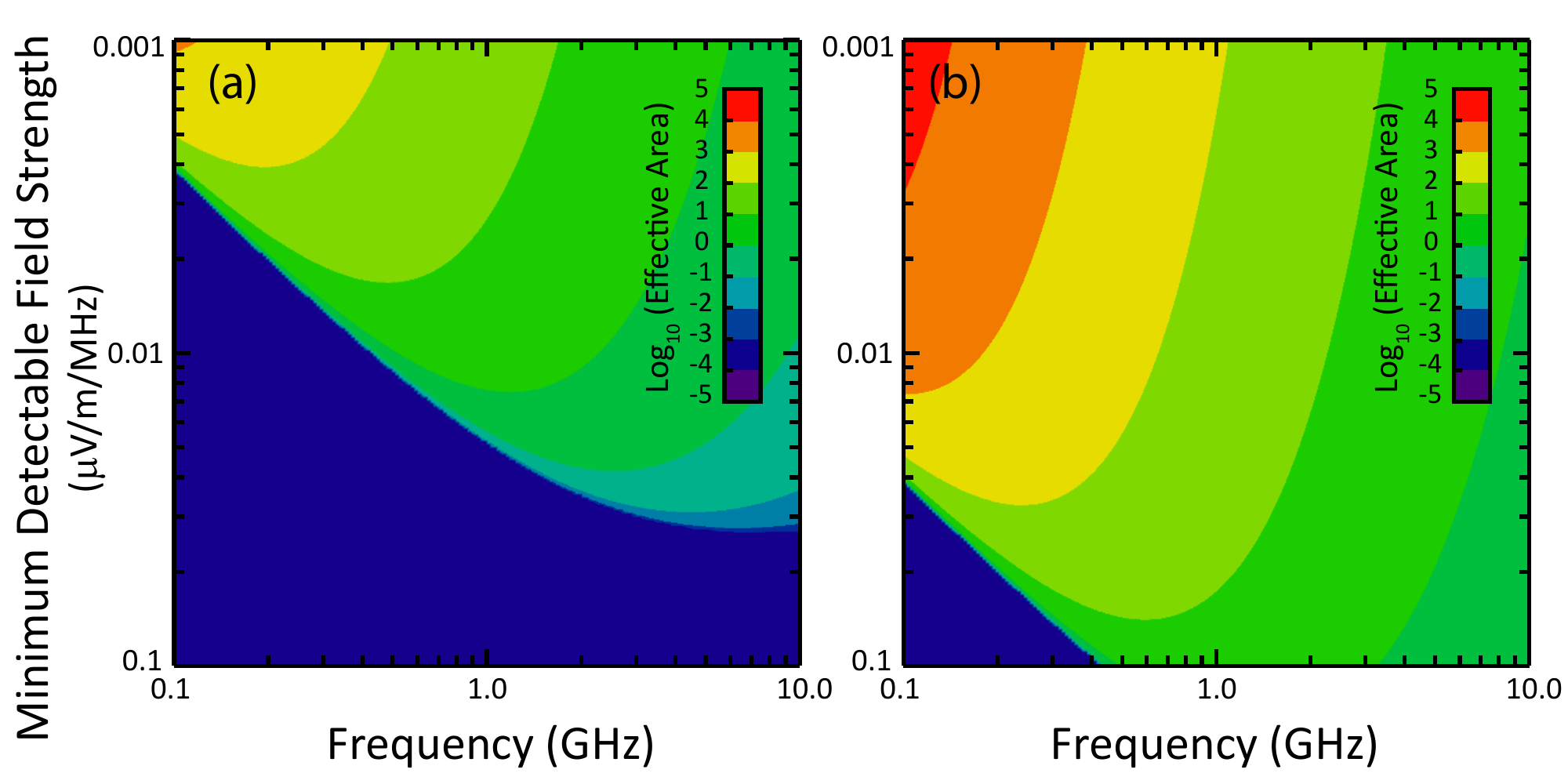}
\caption{$(a)$ Aperture vs. minimum detectable electric field 
and observing frequency for neutrinos with energies 
$E > 10^{21}$ eV. 
$(b)$ Same as panel $(a)$, but for neutrino energies 
$E > 10^{22}$ eV.\label{Fig:Emin-dE}}
\end{center}
\end{figure}

The aperture dependence as a function of both  telescope sensitivity 
($\field_{min}$) and observing frequency is shown as a surface plot 
in Fig.~\ref{Fig:Emin-dE}. 
Panel (a) shows the total aperture for neutrino energies 
exceeding $10^{21}$ eV, while panel (b) shows aperture values 
for $E > 10^{22}$ eV. 
The dark blue regions (zero aperture) correspond to the 
low-$E$ cutoffs seen in Figs.~\ref{Fig:Psi} and \ref{Fig:Emin-dF}. 
Note that for a fixed neutrino energy and telescope 
sensitivity ($\field_{min}$), the maximum aperture results from 
choosing the lowest observing frequency that avoids the
sharp cut-off region.


\subsection{Aperture optimization}

One benefit of scaling laws is the ability to make optimization
calculations using analytic derivatives.
The aperture is energy sensitive, so to maximize the neutrino detection
rate, we must make some assumption about the neutrino energy spectrum.
We will consider two cases: a power-law distribution 
and a mono-energetic spectrum. 

First consider a
canonical power-law neutrino spectrum that scales with neutrino energy
as $E^{-2}$.
We wish to choose frequency $\nu$ to optimize
\beq
\scriptr \ \propto \ \int dE \ E^{-2} P(E) \ .
\eeq
Let us first assume that the surface roughness dominates the aperture,
so we use $\Psi_{dr}$, and note that this will be most appropriate
at higher frequencies where the detectable Cerenkov cone is narrow.
Then
\beq
\scriptr \ \propto \ \int dE \ E^{-2} {\lgamma \over \lnu} f_o^3 \Delta_o \ ,
\eeq
so
\beq
\scriptr \ \propto \ \nu^{-2} \int dE \ E^{-5/3}
\left [ ln \left ({\nu \ E \over 1+(\nu/2.3)^{1.23}}\right )
\ + \ k \right ]^{3/2} \ .
\eeq
We can scale the frequency out of the energy integration using $y=\nu E$,
and neglecting the weak dependence in the $\nu/2.3$ GHz term in the 
logarithm, yields that $\scriptr$ scales nearly with $\nu^{-4/3}$.
Thus the optimal frequency to maximize aperture is as low as possible, at the expense
of increasingly larger neutrino cut-off energy sensitivity and technical issues such as RF interference and ionospheric pulse dispersion.

If, on the other hand, if we are targeting a particular energy
$E_o$, we can focus $\scriptr$ on just that $E_o$, and find
\beq
\scriptr \ \propto \ \nu^{-2} 
\left [ln \left ({\nu \ E \over 1+(\nu/2.3)^{1.23}}\right )
\ + \ k \right ]^{3/2} \ .
\eeq
Evaluating the derivative of this expression, again
neglecting the variation
in the $1+(\nu/2.3)$ term,
leads to the conclusion that the detection rate is optimized when
the frequency is chosen such that
$f_o = \sqrt{3/4} = 0.87$.
This means that for for detection of neutrinos at energy $E_o$,  $f_o$ values substantially above unity
are wasting aperture and would be
better served by reducing $\nu$ (and $f_o$). Likewise,  $f_o$ values
substantially below unity are encountering the energy
cutoff problem, so it is preferable to increase the observing frequency.

At low frequencies surface roughness
no longer dominates, and instead it is the broadly detectable
Cerenkov cones that controls the aperture ($\Psi_{ds}$) for higher
energy neutrinos.
Then we would find for an $E^{-2}$ power law
\beq
\scriptr \ \propto \ \nu^{-3} \int dE \ E^{-5/3}
\left[ln \left ({\nu \ E \over 1+(\nu/2.3)^{1.23}}\right )
\ + \ k \right ]^{2} \ ,
\eeq
which scales roughly like $\nu^{-7/3}$ and favors even
more heavily the lower frequencies.
However, if a particular energy $E_o$ is the target,
optimization occurs near $f_o = \sqrt{2/3} =  0.82$.

In summary, for power-law neutrino spectra with no high-energy cutoff, 
the detection rate is maximized at the lowest possible observing frequency, 
neglecting technical problems with RF interference and ionospheric dispersion. 
However, if a search is optimized for a particular neutrino energy range, 
the aperture (and consequent event rate) is maximized when the 
dimensionless parameter $f_o\sim0.8$. 
From eq. (\ref{fodef}) this condition can be written
\beq
\frac{\field_o}{\field_{min}}= e^{{f_o}^2}\sim1.9
\eeq
We can recast this condition to a more tractable 
form using eq. (\ref{eqn:e0}),
\beq
\nu_{opt}\
{\left[
1+  {  \left(\frac{\nu_{opt}}{2.32}\right)  }^{1.23}
\right]}
^{-1}
= 43.2\ \rm{GHz}\cdot 
{\left[\frac  {E}  {\rm{ZeV}} \right]}^{-1} \ 
\left[\frac {\field_{min}}  { \mu\rm{V\ }\rm{m}^{-1}\rm{\ MHz}^{-1}}\right]\ ,
\eeq
where $\nu_{opt}$ is the observing frequency which 
maximizes aperture at neutrino energy $E$.

\begin{figure}[h]
\begin{center}
\includegraphics[width = 6in]{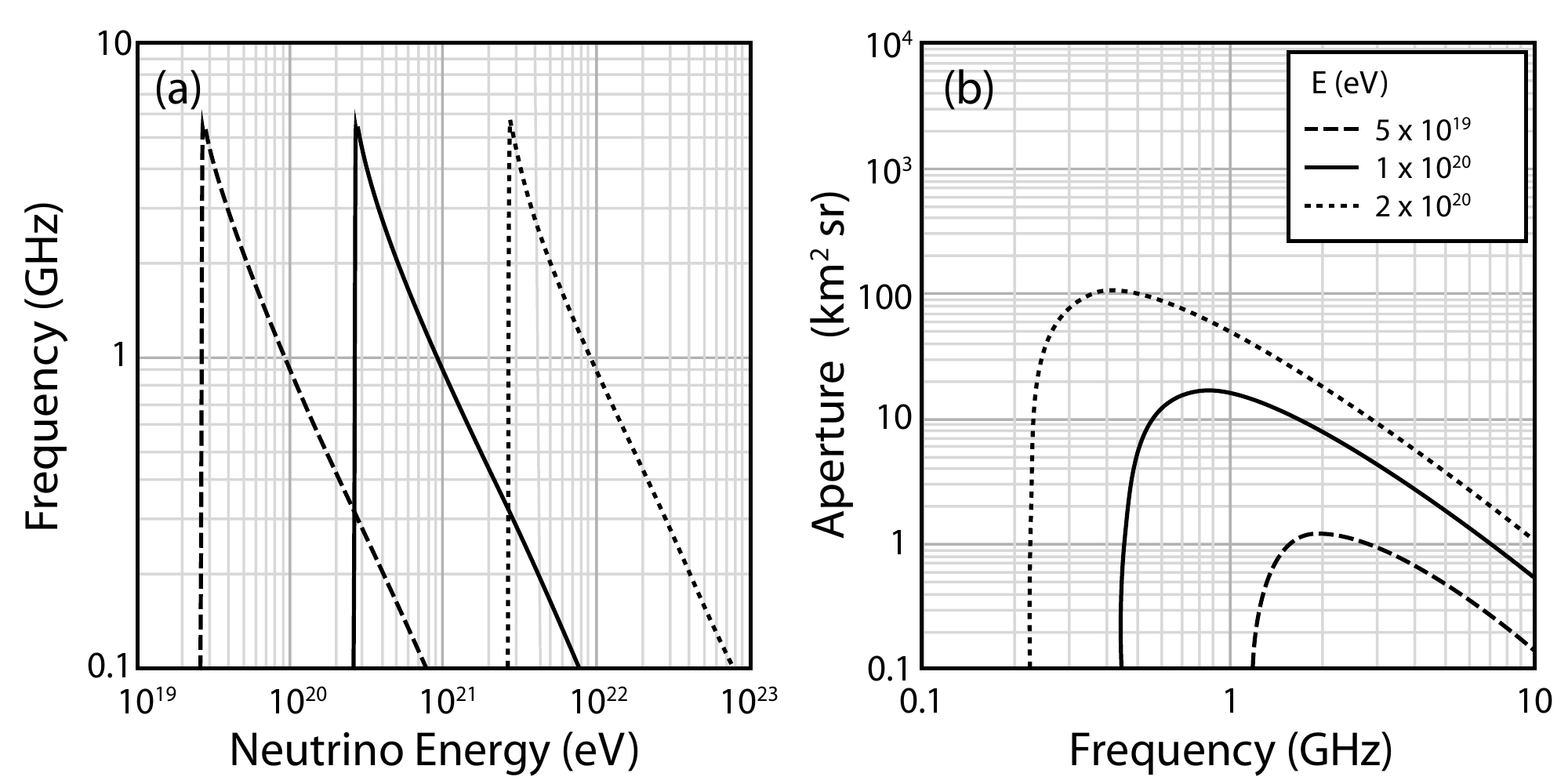}
\caption{(a) Optimal observing frequency vs. target neutrino energy for 
threshold $\field_{min}$ = 0.001 (dashed line), 0.01 (solid line ), 
and 0.1 (dotted line)\edim. 
(b) Aperture vs. observing frequency for $\field_{min} = 0.001$\edim 
and neutrino energy $E = 5\cdot10^{19}$\ eV (dashed line), 
$\ 10^{20}$\ eV (solid line), and $5\cdot10^{20}$\ eV (dotted line). 
\label{Fig:Optimize}}
\end{center}
\end{figure}

Fig.~\ref{Fig:Optimize}(a) shows the frequency which maximizes 
aperture for a given target neutrino energy for threshold 
sensitivities $\field_{min} =$ 0.001, 0.01, and 0.1\edim. 
Fig.~\ref{Fig:Optimize}(b) plots the aperture as a function of 
frequency for a threshold sensitivity $\field_{min} =$ 0.001\edim for 
three target energies near the GZK cutoff.
As discussed in section \ref{min-efield}, this sensitivity corresponds 
to much larger aperture (e.g., SKA)  than current searches. 
Note that the optimal search frequency for GZK neutrinos depends 
strongly on the exact model for the cosmogenic neutrino energy spectrum, 
which varies with UHECR source model \citep[e.g.,][]{Kalashev:2002}.

\subsection{Telescope requirements for GZK-neutrino searches}

In order to probe near GZK energies ($E\ltwig10^{20}$\ eV), 
the minimum detectable electric field must be  
$\field_{min}\ltwig10^{-3}$ \edim, 
largely independent of frequency. 
From Fig.~\ref{Fig:emin-plot} we can see that even 
for very wide bandwidths, this requires an effective collecting 
area exceeding 50,000\ {m}$^2$, approximately the area 
of the Arecibo telescope.  
Since most detection schemes require coincidence on multiple 
telescopes to discriminate against accidental pulses, 
an array of multiple Arecibo-class telescopes are required. 
Current experiments using multiple 10-100 m diameter telescopes 
can only probe neutrino energies well above GZK, where more 
exotic neutrino production processes may be operating 
\citep[e.g.,][]{Kalashev:2002}. 
We therefore conclude that only next-generation arrays, e.g.,  
the full SKA ($A_e\sim10^6$ m$^2$, $\field_{min}\sim10^{-4}$ \edim) 
will have sufficient collecting area to probe GZK energies. 
However, even the SKA will  have an aperture of only a few km$^2$ at 
the GZK cutoff $E_{GZK}\sim10^{19.6}$ eV. 
Given an expected GZK neutrino flux 
$F_{\nu}\sim$1 km\mm\ yr\mm\ sr\mm\  
\citep[e.g.,][]{Gusev:2006}, the event rate might 
only be a few counts per observing month.
Hence an even higher energy neutrino population, if it
exists, might continue
to prove easier to detect, even with SKA-class technology.

\subsection{Comparison with Monte Carlo simulations}

Finally, we address the question of how the present analytic 
calculations compare with previously published Monte Carlo simulations 
\citep[e.g.,][]{Gorham:2001, Beresnyak:2003, Williams:2004, James:2009}. 
In general, a direct comparison with each simulation is problematic, 
since the input physics model (neutrino and radio extinction lengths, 
Cerenkov cone width and peak electric field, lunar roughness, detailed 
regolith properties) is significantly different for each simulation. 
However, since we have largely adopted the input physics parameters 
from \citet{James:2009},  a direct comparison is warranted in this case. 

Table~1 shows a comparison of calculated apertures using the analytic 
approximation (equation \ref{aperapprox}) compared with apertures 
reported by \cite{James:2009} in their Fig. 6a (no sub-regolith case, 
different pointings summed),  at a fixed neutrino 
energy $E =$ 1 Zev ($10^{21}$ eV) for convenience. 
For completeness, we also list the upper limit (90\% confidence) 
to the commonly plotted quantity $F(E) = E^2I(E)$, 
where $I(E)$ is the differential neutrino flux,  using the 
``model independent'' expression \citep{Lehtinen:2004}
\beq
F(E)  <  \frac{2.3 \ E} {t_{obs}A(E)}\ ,
\eeq
where $t_{obs}$ is the total observing time. 

Table~1 lists observational parameters, apertures, and flux limits for 
three searches: Parkes \citep{Hankins:1996, James:2007b}, 
GLUE \citep{Gorham:2004}, and Kalyazin \citep{Beresnyak:2005}. 
The table lists the relevant input parameters used in the calculation, 
which were derived from the literature: 
threshold  electric field ($\field_{min}$), center observing 
frequency ($\nu$), and fraction of the Moon's limb sampled ($\zeta$). 
This fraction is difficult to estimate accurately since the searches 
often used pointings with roughly defined beam positions 
(e.g., `limb', `half-limb') and differing beamwidths for pulse 
coincidence schemes with multiple telescopes or frequencies. 
Nevertheless, we made our best estimate based on the experimental 
descriptions, and multiplied the calculated aperture by this fraction, 
since the calculation assumes 100\% limb coverage. 
The columns in Table~1 are: (1) experiment name, (2) threshold electric 
field ($\field_{min}$), (3) mid-observing frequency ($\nu$), (4) total 
observing time ($t_{obs}$), and (5) fractional lunar limb coverage ($\zeta$). 
Columns 6-7 are apertures calculated using eq. (\ref{aperapprox}) ($A_{gmj}$),
and from Fig. 6a of \cite{James:2009} ($A_{jp}$). 
Columns 8-10 are upper limits to the quantity $F(E)$ using aperture 
$A_{gmj}$,  $A_{jp}$, and directly from the original search papers.

\begin{figure}[h!]
\begin{center}
\includegraphics[width = 6in]{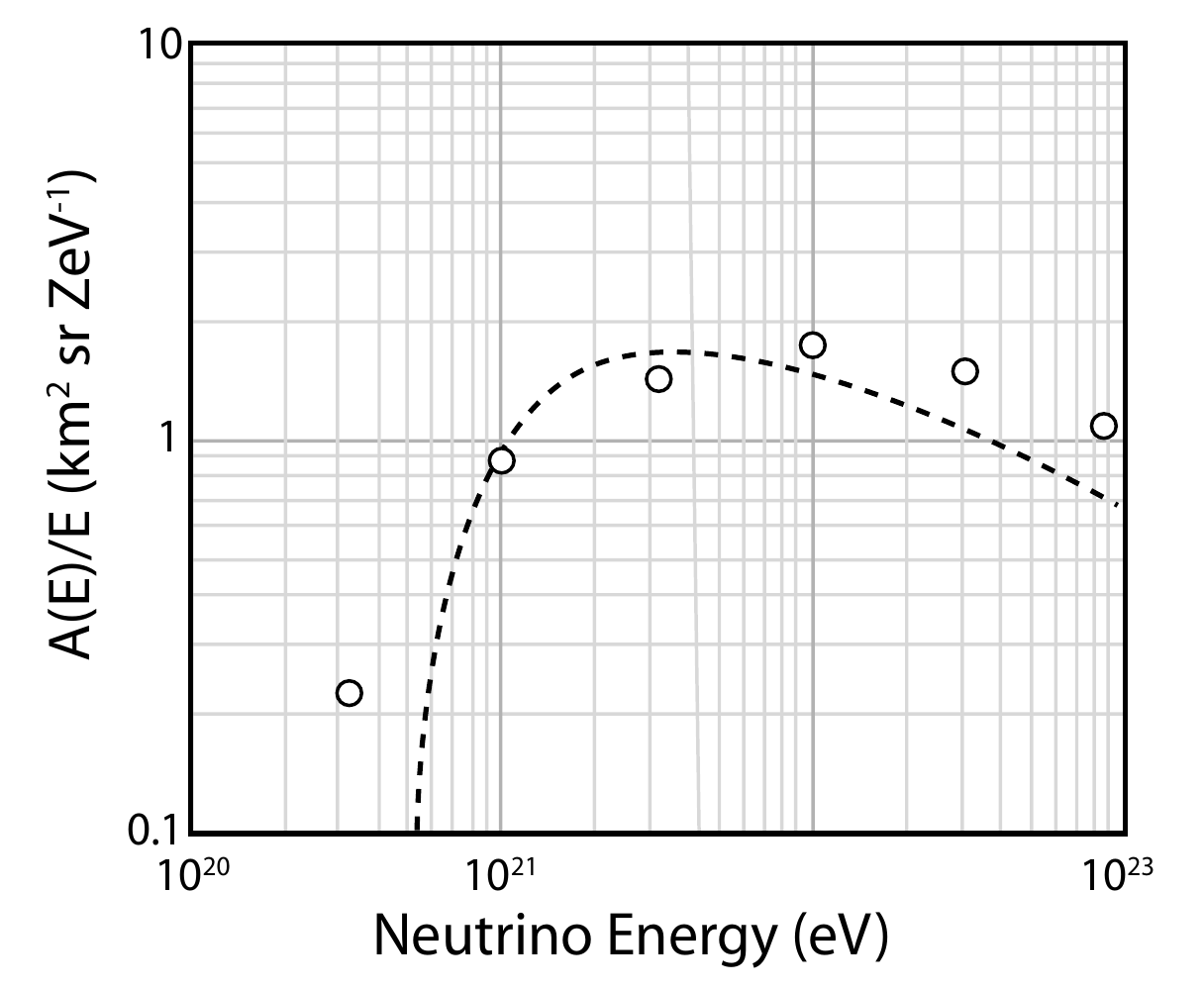}
\caption{Effective aperture divided by neutrino energy versus energy for GLUE experiment \citep{Gorham:2004} calculated using equation \ref{aperapprox} (dotted line) and Monte Carlo calculation of \cite{James:2009} for regolith only (open circles, from their Fig. 6a).
\label{fig:compare}}
\end{center}
\end{figure}

Inspection of Table~1 shows that the apertures calculated at $E=1$~ZeV by the 
analytic approximation 
agree very well with the Monte Carlo results 
of \cite{James:2009} for all three experiments. 
In Fig. \ref{fig:compare} we show a comparison of the ratio of effective aperture to energy as a function of neutrino energy for the GLUE experiment using our analytic calculation (eqn. \ref{aperapprox}) with  \citet[][regolith only]{James:2009}. The agreement is quite good, although there appears to be a small systematic difference with energy scaling.
It is difficult to assess whether or not any of the disagreement could
be traced to weakness in the Monte Carlo simulations, as details of such
calculations are not generally reported with sufficient completeness to 
reproduce the results in detail.
Nevertheless, the overall  conclusion is
that the resulting estimates of neutrino 
flux upper limits are in good agreement, and our results help confirm the 
assertion of \cite{James:2009} that the widely referenced 
GLUE upper limit \citep{Gorham:2004} should actually be an order of 
magnitude higher than previously reported. 

\begin{table}[h]
\centering
\label{table:aperture}
\caption{Aperture, neutrino flux limit comparison at 
neutrino energy $E$ = 1 ZeV}
\vspace{0.3in}
\begin{minipage}{20cm}
\begin{tabular}{lccccccccc}
Experiment & $\field_{min}$ & $\nu$ & $t_{obs}$ & $\zeta$ & $A_{gmj}$ & $A_{jp}$ &log($F_{gmj})$& log($F_{jp})$ &  log($F_{orig})$ \\
 & Vm\mm\ MHz\mm & GHz & Hr &  &\multicolumn{2}{c}{(km$^2$-sr)} & 
  \multicolumn{3}{c}{(GeV\ cm$^{-2}$\ s$^{-1}$\ sr$^{-1}$)}   \\
 \hline
 Parkes (limb) & 0.013 & 1.5 & 2.0 & 0.20 & 2.1 & 2.0 & -1.8 & -1.8 &
 -2.1\footnote{\cite{James:2007} - \cite{Hankins:1996} does not give a flux limit at $10^{21}$ eV}\\
 GLUE (limb) & 0.011 & 2.2 & 
 120\footnote{Upper limit: \cite{Gorham:2004} does not state how many hours were pointed on limb.} & 0.17 & 1.1 & 1.0 & -3.4 &-3.2 & -4.2 \\
 Kalyazin & 0.013 & 2.3 & 31 & 0.11 & 0.7 & 0.6 & -2.5 & -2.3 & -2.2 \\
 \hline\vspace{0.15in}
\end{tabular}
 \end{minipage}
\end{table}

\section{Summary and Conclusions}

Accurate aperture calculations are complicated and laborious and historically have required Monte Carlo simulations.
We have shown that an analytic calculation with a series of simplifying approximations
can reproduce a result comparable  to Monte Carlo 
simulations, and in addition generate simple scaling laws with a straightforward
conceptual interpretation (eq. [\ref{aperapprox}]).
We find that it is crucial to account for 
surface roughness
when apertures are intrinsically low, such as when using
high frequencies from Earthlike distances,
but when the aperture is intrinsically high, such as at low frequencies 
with high-energy neutrinos, or for lunar orbiters, then surface roughness is
of lesser significance.
We also find (Fig. [\ref{Fig:Psi}]) 
that downward neutrinos significantly dominate over upward
neutrinos for higher
neutrino energies, and this is 
especially true for lower frequency observations
($\nu \ltwig 300$ MHz).
However, 
at the energies nearest to the
GZK regime, which is perhaps of greatest cosmological
importance, both upward and downward neutrinos may contribute comparably to
the detection rate.

These conclusions imply that the first detected GZK neutrino, if detected
at high frequency to give a more favorable energy cutoff,  may be a
downward neutrino whose Cerenkov signal will have escaped by virtue of a lunar 
roughness feature, or it may with similar likelihood
be an upward neutrino that will have made it through a
long column of lunar rock and then created an upward Cerenkov flash that
did not require assistance from lunar roughness.
Alternatively, if the first GZK neutrino
detection comes at lower frequency, it will 
likely be
because a downward neutrino sent out a very weak radio signal that did
not require lunar surface roughness to escape, but did require an
extraordinarily sensitive radio instrument to detect.
It is also possible that the first UHE neutrino will come from pion decay
near the acceleration region of a UHE proton, and then it could come
at an energy well beyond the GZK regime.
In that case, the detection could come at low or high frequency, depending
on whichever instrument first achieves the necessary sensitivity.

Another important conclusion
is that simple scaling laws, with all their
extreme portability, may be used to optimize 
experimental design within a wide range of constraints.
They are also useful for making instant comparisons between the apertures of
past experiments using a standardized treatment of
the relevant input parameters.
In this way, the impact of the various inconsistencies between models,
such as the detectable Cerenkov cone width,
the treatment of transmission and solid-angle magnification,
and the role of surface roughness, 
can be addressed without having to run Monte Carlo simulations with
exactly the same parameters.
Indeed, uncertainties in the details of such simulations make it difficult
for us to 
resolve several inconsistencies
between our results and others quoted in the literature, whether they be
due to problems in the Monte Carlo simulations or in our own analytic
approximations, especially our use of small-angle approximations and a
constant transmission coefficient. For example, we have not been able to determine the source of the order of magnitude increase in the neutrino flux upper limit compared with the published GLUE upper limits (cf. Table 1). This discrepancy was also reported by \cite{James:2009} using a Monte Carlo simulation.

One particularly robust result we obtain is that to optimize the detection
of neutrinos at a given energy, one should choose a frequency that will
yield a value for the $f_o$ parameter 
of roughly 0.8, which implies a  
ratio of the maximum to the minimum detectable field of
about 2,  for that energy.
Inverting this implies that past experiments at given frequencies 
are best tailored to the neutrino energies
for which the maximum field generated in the telescope by neutrinos
of that energy is about twice the 
minimum detectable field for that instrument.

We would like to acknowledge helpful discussions with Hallsie Reno, John Ralston,
and Clancy James, who were especially helpful guides in areas of this effort 
outside our own expertise.

\bibliography{uhe-neutrinos}

\section{Appendix A: The effective neutrino extinction pathlength}

When UHE neutrinos strike the Moon, 
there is a roughly 1/3 chance \citep{Reno:2005}
that they will initiate via the neutral current interaction a hadronic
cascade that will leave roughly 80\% of the neutrino energy still in the
highly forward-scattered neutrino.
The remaining 2/3 of the time, a charged current interaction will destroy
the neutrino, but still result in the same roughly 20\% of the energy
going into the hadronic cascades we wish to detect.
Thus from the point of view of detecting the cascade, we do not care whether
the interaction was via neutral or charged current.
However, this is a relevant
question when considering whether or not the neutrino survives
the encounter and can initiate further cascades.

Since the observed radio signals are from a narrow layer at
the surface, it is highly unlikely that the same neutrino could be
involved in multiple detectable showers.
Nevertheless, some of the neutrinos we are hoping to detect (labeled 
the ``upward'' neutrinos) may have to
make their way through a considerable amount of lunar material before
arriving at their observable shower location.
For these neutrinos, it does matter if they can survive prior neutral-current
interactions, losing only $\sim 20$\% of their energy each time.
For example, a neutrino that suffers two neutral-current interactions and
no charge-current interactions can arrive in the domain of interest with
roughly 64\% of its energy intact, so at the energy of interest, this just
means that there can be a contribution from initially higher energy
neutrinos.
This ``downgrading'' effect is easily accounted for by increasing the 
effective neutrino extinction path over and above $L_{\nu}(E)$,
by a factor that depends on the
steepness of the energy spectrum, thereby increasing
the number of ``upward'' neutrinos that contribute to the aperture
at any given energy.
Once this has been accounted for, the
actual reaction rates for a given neutrino population
are still proportional to $L_{\nu}^{-1}$, so the effect is not the
same as a global correction to $L_{\nu}$, it is merely an effective
reduction in neutrino extinction at energy $E$.

Here we derive in the simplest case
the enhancement in the effective extinction
pathlength, so we find the factor $\psi$
we use to multiply $\lnu(E)$ in eqs. (\ref{tauexpression}) and 
(\ref{alphaodef}),
for a power-law neutrino energy distribution, 
with flux per unit energy 
$I(E) \propto E^{-2}$.
The power-law nature of the problem imposes a scale invariance that
allows us to seek a {\it constant} factor by which all effective extinction
lengths are altered, where the extinction of $I(E)$ must be modified
by the appearance in $I(E)$ of downgraded neutrinos from initially
higher energies.
This results in an 
{\it effective} extinction coefficient ${\bar \chi(E)}$, differing
from the actual extinction coefficient $\chi(E)$,
such that $I(s) = I_o e^{-{\bar \chi}s}$.
The net extinction over length $ds$ must obey
\beq
{dI \over ds} \ = \ -\chi(E) I_o e^{-{\bar \chi}s} \ + \ 
b\cdot1.077\cdot0.8\cdot\chi(E) I_o e^{-{\bar \chi}s} \ ,
\eeq
where $b$ is the branching ratio for the survival of the neutrino, which
here is
$b = 1/3$ \citep{Cooper:2008}, 
1.077 comes from assuming that $\chi(E) \propto
E^{1/3}$ so the $\chi$ for the downgraded neutrinos is slightly
higher than $\chi(E)$, and the 0.8 factor comes from the combination of
the $E^{-2}$ neutrino power law (assumed only for simplicity here) and
the fact that higher energy bins are squeezed into narrower
energy bins when 20\% of the neutrino energy is lost to the hadronic
shower.

It remains only to note that the above equation takes on the desired form
\beq
{dI \over ds} \ = \ -{\bar \chi}  I_o e^{-{\bar \chi}s}
\eeq
when
${\bar \chi} = \chi/1.4$.
Hence, the extinction of neutrinos with $I(E) \propto E^{-2}$
acts as though the extinction length
was larger by a factor $\psi = 1.4$.
Had we instead used a power
law of -2.7 instead of -2, the factor would have been 1.3, but the factor
could be higher for flatter spectra, such as in the rising hump of
the GZK cutoff region.
It is even possible for the neutrino flux at a given energy at the start of
the GZK hump to experience a region of neutrino {\it enhancement} inside
the Moon, but we do not deal with this possibility here because no
suitably general assumptions about the shape of the neutrino spectrum can be
applied at this time.
If GZK neutrinos are detected, they will likely come at energies
near the peak or the
falling part of the GZK hump, so we are probably not underestimating
the neutrino population that is being passed down from higher energies
as they cross lunar rock.

\end{document}